# Modelling Threat Causation for Religiosity and Nationalism in Europe

Dr. Josh Bullock, Kingston University, London
Dr. Justin E. Lane, ALAN Analytics; Slovakia, NORCE Center for Modeling Social Systems, Kristiansand, Norway
Dr. Igor Mikloušić, Pilar Institute, Zagreb, Croatia
Professor F. LeRon Shults, University of Agder, Norway; NORCE Center for Modeling Social Systems, Kristiansand, Norway

## Introduction

Europe's contemporary political landscape has been shaped by massive shifts in recent decades caused by geopolitical upheavals such as Brexit and now, COVID-19. The way in which policy makers respond to the current pandemic could have large effects on how the world looks after the pandemic subsides. A region that has been defined by liberalism and freedom for the better part of a century has experienced a rise in nationalism and populism that has not been seen since after World War I. Historically, limitations on individual freedoms and new waves of legislation impeding freedom have accompanied the rise of nationalism and populism in Europe. These developments have also been accompanied by increased xenophobia and out-group distrust, which have negative effects on economic cooperation globally and economic mobility locally. Therefore, understanding what motivates and drives populist and nationalist movements from a psychological and social perspective is critical to promoting policies that can promote individual freedom in Europe.

There are still significant gaps in the scholarly literature on populism and nationalism, and psychological studies all too often neglect the evolutionary basis for human psychological tendencies and motivations that are necessary for nationalist and populist movements. In particular, there is a lack of attention to the role of evolved human psychology in responding to persistent threats, which can fall into four broad categories in the literature: predation (threats to one's life via being eaten or killed in some other way), contagion (threats to one's life via physical infection), natural (threats to one's life via natural disasters), and social (threats to one's life by destroying social standing; exile from a group would have been equivalent to a death penalty in ancient ancestral environments) (Liénard & Boyer, 2006;; Shults, Gore, et al., 2018). These threats have been discussed in light of their effects on religion and other forms of behaviour, but they have not been employed to study nationalist and populist behaviours.

In what follows, two studies are presented that begin to fill this gap in the literature. The first is a survey used to inform our theoretical framework and explore the different possible relationships in an online sample. The second is a study of a computer simulation. Both studies described below (completed in 2020) found very clear effects among the relevant variables, enabling us to identify trends that require further explanation and call for additional research as we move toward models that can adequately inform policy discussions.

**Theoretical Background**

Although many theories from a variety of disciplines bear on the relationships between threat and populism/nationalism studied in this paper, two sets of sometimes overlapping literatures are particularly important for our purposes here. The first is *identity fusion theory* (Bonin & Lane, n.d.; Jong et al., 2016; Swann et al., 2012; Whitehouse et al., 2017; Whitehouse & Lanman, 2014), which is closely linked to the concept of "sacred values". Gómez (2020) points out that the current challenge of the Covid-19 pandemic is converting some citizens into "devoted actors" and increasing fusion with various groups. When contagion threats strengthen nationalism, this intensified sense of unity may make an in-group stronger, but it also poses a number of potential problems for out-group dynamics and intergroup conflict. Gomez argues these changes may include "denial of the group's wrongdoings… willingness to participate in extreme forms of protest on behalf of the group; maximizing the ingroup's advantage over the outgroup even at one's personal expense; protecting the group's reputation… relative intergroup formidability… and the desire to retaliate against outgroup members" (p.2-3). All of this contributes to out-group hostility. In the US context, culpability has sometimes been consigned to the Chinese with major media outlets and government officials, including the president, repeatedly using the term "Chinese virus", which further entrenches the notion that the pandemic is the fault of "others." This perpetuates the idea that "foreigners are also associated with semantic concepts that connote disease" (Faulkner et al. 2004: p.333).

A second body of literature that is relevant for our current purposes is the literature on *moral foundations theory*. The basic idea here is that human morality evolved with five distinct "foundations," which are differentially distributed within human populations. Some combination of these five emotionally charged moral foundations form the basis of each individual's normative preferences: care/harm, fairness/cheating, loyalty/betrayal, authority/subversion, and purity/degradation. The first two foundations are sometimes referred to as "individualizing," while the latter three are considered "binding." Liberals tend to be guided primarily by the first two of these dyads, suppressing the other three, while conservatives rely on all five when making moral judgments (Alizadeh et al., 2019; Graham et al., 2009; Iyer et al., 2012). Previous research testing this theory suggests that "individualizing" moral foundations can have a negative relationship with negative behavioural intentions (and a positive relation with positive intentions), while "binding" moral foundations are correlated oppositely. This is relevant for our purposes because both of our studies were motivated by our interest in discovering the causal interactions among threats (of the four types mentioned above), religiosity, nationalism, and intergroup conflict. However, our evolutionary framework focuses on the role of threat perception, rather than moral belief commitments, as a factor in differences between poles in the political spectrum.

**Study 1**

Study 1 collected data on the relationships between nationalism, religiosity, national and religious identification, threat perception, and sentiment toward different groups (using the same measures as the World Values Survey), particularly focusing on immigrants. Data was also collected on social media use and consumption of TV based media. Participants were also asked about their experience during the Covid-19 pandemic and their infection status.

**Methods**

An online survey was developed and deployed online with SurveyGizmo. Participants (N=2000) were recruited online through posting on forums and sharing in online social networks. Most participants were recruited on MTurk.

**Measures**

There were many survey measures used to assess our key variables. These are described here:

*Covid-19 infection status*
Participants were asked if they had tested negative for Covid-19, if they had symptoms and weren't tested, if they currently had it, and if they had recovered.

*Nationalism*

We utilized an adapted version of the Nationalism Scale first put forward by Mansillo (2016). The key addition was an additional item with the prompt "to be truly part of my nation, one must be a specific race or ethnicity". We found the original scale items to have acceptable reliability ($\alpha = .88$; 95%CI = [.87;.89]), as well as the extended scale with our additional item ($\alpha = .90$; 95%CI = [.89;.90]).

*Social Identification*

Participants were asked to state their nationality. They were then prompted to complete the Postmes 4-item social identification scale where the target group was their nation (Postmes et al., 2005). We found the scale items to have acceptable reliability ($\alpha = .92$; 95%CI = [.91;.93]). Participants were asked to state their religious affiliation. They were then prompted to complete the Postmes 4-item social identification scale where the target group was their religious group. We found the scale items to have acceptable reliability ($\alpha = .94$; 95%CI = [.93;.95]).

*Fusion*

Individuals were also asked to state their nationality. They were then prompted to complete the verbal fusion scale where the target group was their nation (Gómez et al., 2011). We found the scale items to have acceptable reliability ($\alpha = .95$; 95%CI = [.945;.953]). Participants were also asked to state their religious affiliation. They were then prompted to complete the verbal fusion scale where the target group was their religious group. We found the scale items to have acceptable reliability ($\alpha = .96$; 95%CI = [.95;.96]).

*Supernatural Belief*

To measure supernatural beliefs, an adapted version of the Supernatural Belief Scale (Jong et al., 2013) was used. The amendment involved the inclusion of two additional items. The first presented the prompt "There exists a universal force of justice that you could call karma". The second was that "The universe has the ability to affect the events in our lives and provide balance." We found the original scale items to have acceptable reliability ($\alpha = .98$; 95%CI = [.975;.978]), as well as the extended scale with our additional item ($\alpha = .98$; 95%CI = [.975;.978]).

*Threat*

Threat was assessed with a scale that we devised in our team to measure 4 key evolutionary threats (predation, contagion, social, and natural hazards). We added the additional dimension of financial threats because of the likelihood that participants' key concern during the Covid-19 pandemic could be the loss of their job. All items were based on a 5 item Likert scale (strongly disagree to strongly agree). The items were as follows:

1. I feel financially secure
2. I worry about my financial situation
3. I am concerned about natural disasters
4. Natural disasters are not a threat
5. I am secure in my social standing
6. My social standing is threatened
7. I will contract COVID-19 (coronavirus) or another ailment
8. I am confident that I will not contract COVID-19 (coronavirus) or another ailment
9. I worry that I might be killed by another person (or animal)
10. I am confident that I will not be killed by another person (or animal)

In this measure, item 1, 4, 5, 8, and 10 are reverse coded. Items 1, 2 address financial threats. Items 3, 4 address natural threats. Items 5, 6 address social threats. Items 7, 8 address contagion threats. Items 9, 10 address predation threats. There were two additional threat questions that were asked but not included in the threat measures.

*Our way of life is under threat because of COVID-19 (coronavirus)*
*Our way of life is threatened because of the number of foreigners or immigrants*
*I believe the world will never be the same because of COVID-19 (coronavirus)*

We found that the scale items addressing the 4 key threats (predation, natural, social, contagion) did *not* have acceptable reliability ($\alpha$ = .42; 95%CI = [.37;.46]), and that adding the financial threat measure did not change this conclusion ($\alpha$ = .55; 95%CI = [.51;.58]). To investigate whether there is a multidimensional structure to this scale we analysed the dimensionality of the scale. Using factor analysis, we found that there are multiple dimensions to the measure. A scree plot analysis suggests that 4 dimensions are optimal.

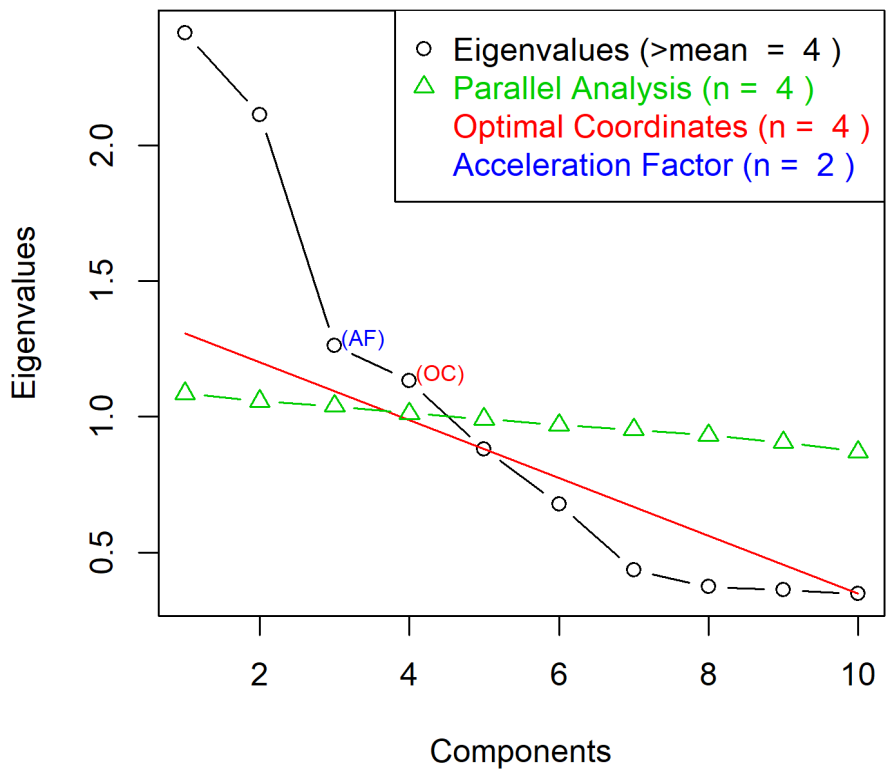

A mapping of the principle components analysis suggests that there is a common directionality to the items (after the appropriate reverse coding).

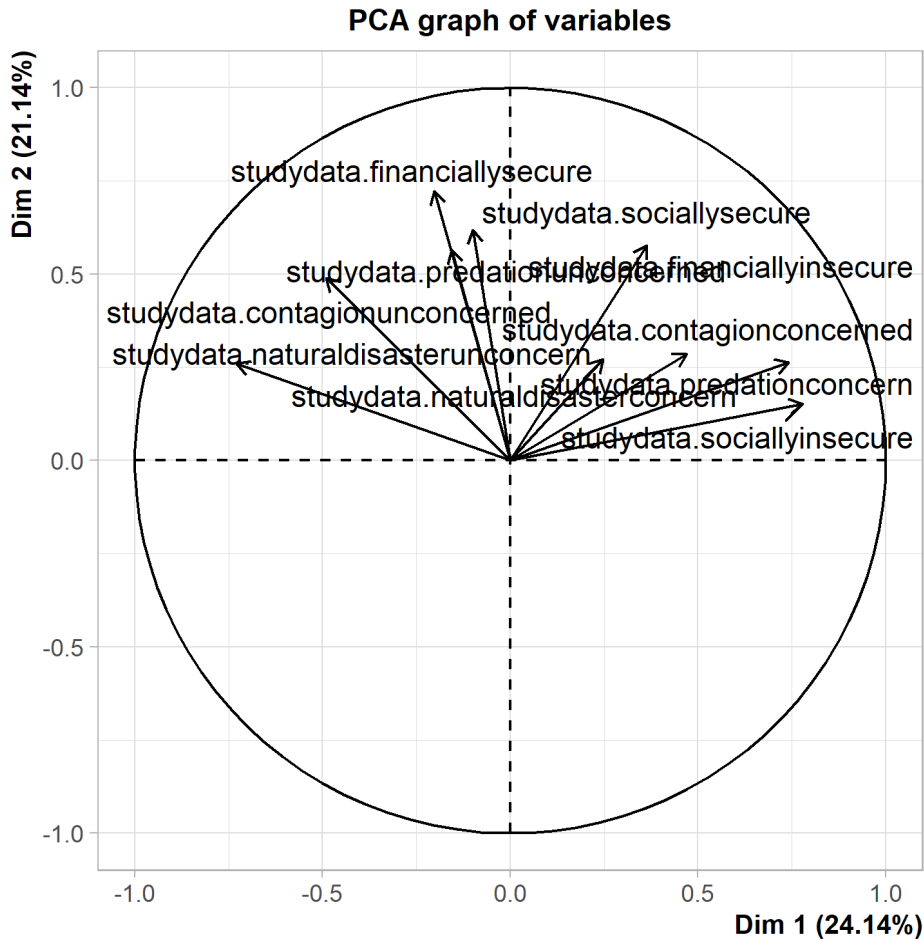

As such, later analyses will treat the treat measure as a multi-component scale, not a single dimensional scale.

*Personality*
We used a standard 10-item Big-5 measure. We tested whether there was a significant fit for the 5-factor model and did find a significant fit (.78) using varimax rotation. The resulting root mean square of the residuals was found to be 0.09 (chi squared = 1516.21 $p<.01$). As such we utilize the 5-factor measure as intended.

*Religious attendance*
Participants were asked how frequently they attend religious services using a 10-item ordinal scale from daily to never.

*Social media use*
Participants were asked how much time they spend on social media (from rarely, to more than 2 hours a day).

*Brexit*
Participants from the UK were asked how they voted in the 2016 EU referendum and if they still support that decision.

*Politics*
Participants were first asked to report how liberal or conservative they are on economic issues using a visual slider anchored at very liberal and very conservative.
They were then asked to report how liberal or conservative they are on social issues using a visual slider anchored at very liberal and very conservative.
Lastly, they were asked what party they voted for in the last election, if any.

*Outgroups*
Using a Likert scale from very negative to very positive, participants were prompted to answer how they feel toward the following groups: Jews, Immigrants, Atheists, people of other races than their own, Christians, and Muslims. We also used questions taken from the world values survey which prompted the participant with "On this list are various groups of people. Could you please mark any that you would not like to have as neighbours?" Participants are then presented with the options: "drug addicts; people of a different race; people who have aids; immigrants/foreign workers; homosexuals; people of a different religion; heavy drinkers; unmarried couples living together; people who speak a different language".

*Media use*
Participants were asked how frequently they watch or read the news online (from less than once a day to 3+ times a day)

*News Sources*
Participants were also asked to list the news sources they follow.

**Participants**
After removing invalid responses, we were left with N=2,018 participants. Invalid responses were deemed to be those that were too incomplete for use or did not follow directions.
The sample had 1038 males, 970 females, and 10 other.

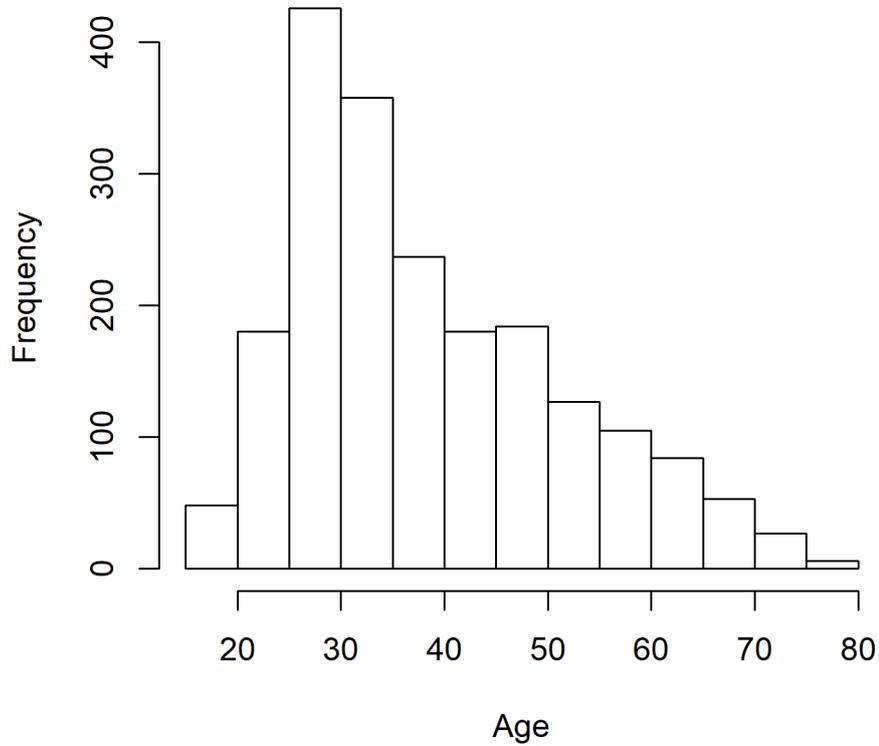

**Participant Age Distribution**

Participants had a mean age of 39.06 (*sd* = 13.00). Their age distributions are visualized above.

**Results**

*Religious Demographics*

Generally, we found that we have oversampled atheists, with far more people reporting that they do not attend religious services than expected.

## Religious Attendance

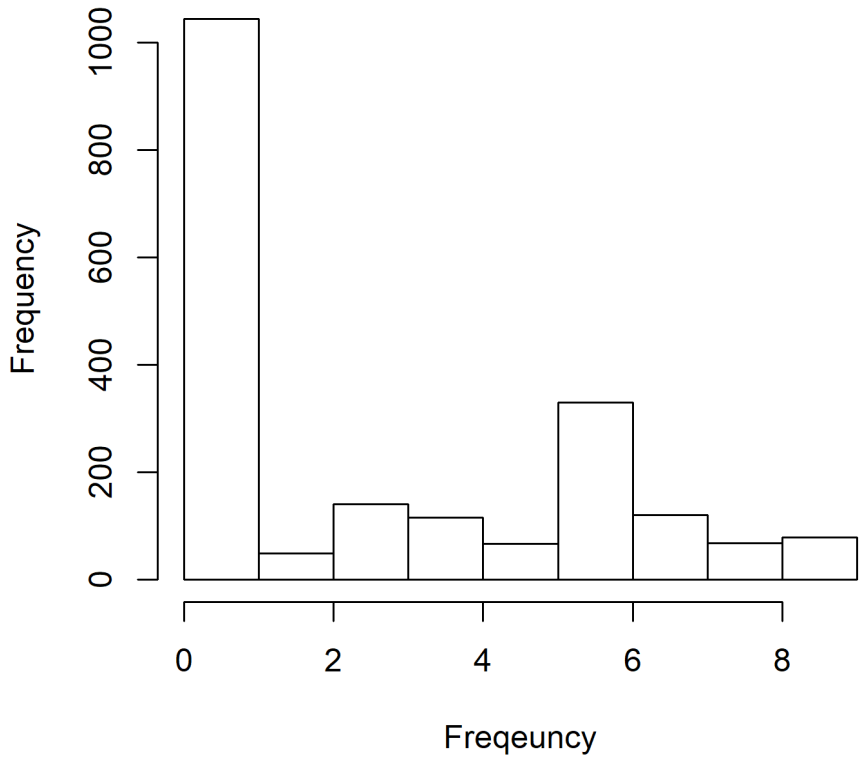

This distribution is consistent with the categorical self-identification responses:

| Religion | Frequency |
|---|---|
| Agnostic | 169 |
| Atheist | 485 |
| Buddhist | 33 |
| Catholic | 481 |
| Church of England/Anglican | 37 |
| Evangelical/Pentecostal/Charismatic | 40 |
| Hindu | 64 |
| Humanist | 70 |
| Jewish | 26 |
| Muslim | 35 |
| None | 174 |
| Other | 39 |
| Protestant (misc.) | 246 |
| Spiritual but not religious | 119 |

Responses on the supernatural belief scale, however, suggested that supernatural belief was still strong. Nevertheless, the clear signature of non-belief can be readily observed, and the addition of our two amended items did not change this distribution.

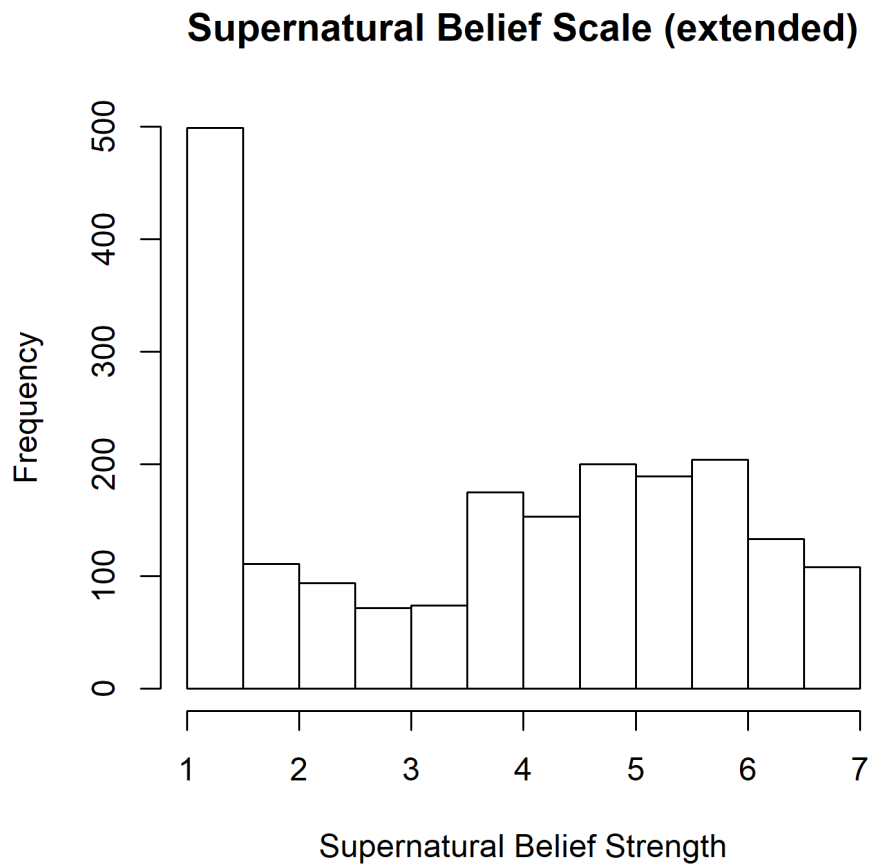

**Supernatural Belief Scale (extended)**

In addition to the standard supernatural belief scale, we added two items for karma and universal force (described in the measures section). The variation of support for those two beliefs was interesting because it reveals that these beliefs are held by individuals who are otherwise "nonreligious".

Regarding the idea of a universal force there was general acceptance but significant differences between groups ($F(13,2000) = 89.47, p < .01$)). Atheists had a long-tail distribution ranging from no support to a tailing off of support, but we see fairly normal distributions in this belief for Agnostics, Humanists and nones. The Spiritual but not Religious have a similar signature to Protestants, trending toward acceptance of the idea.

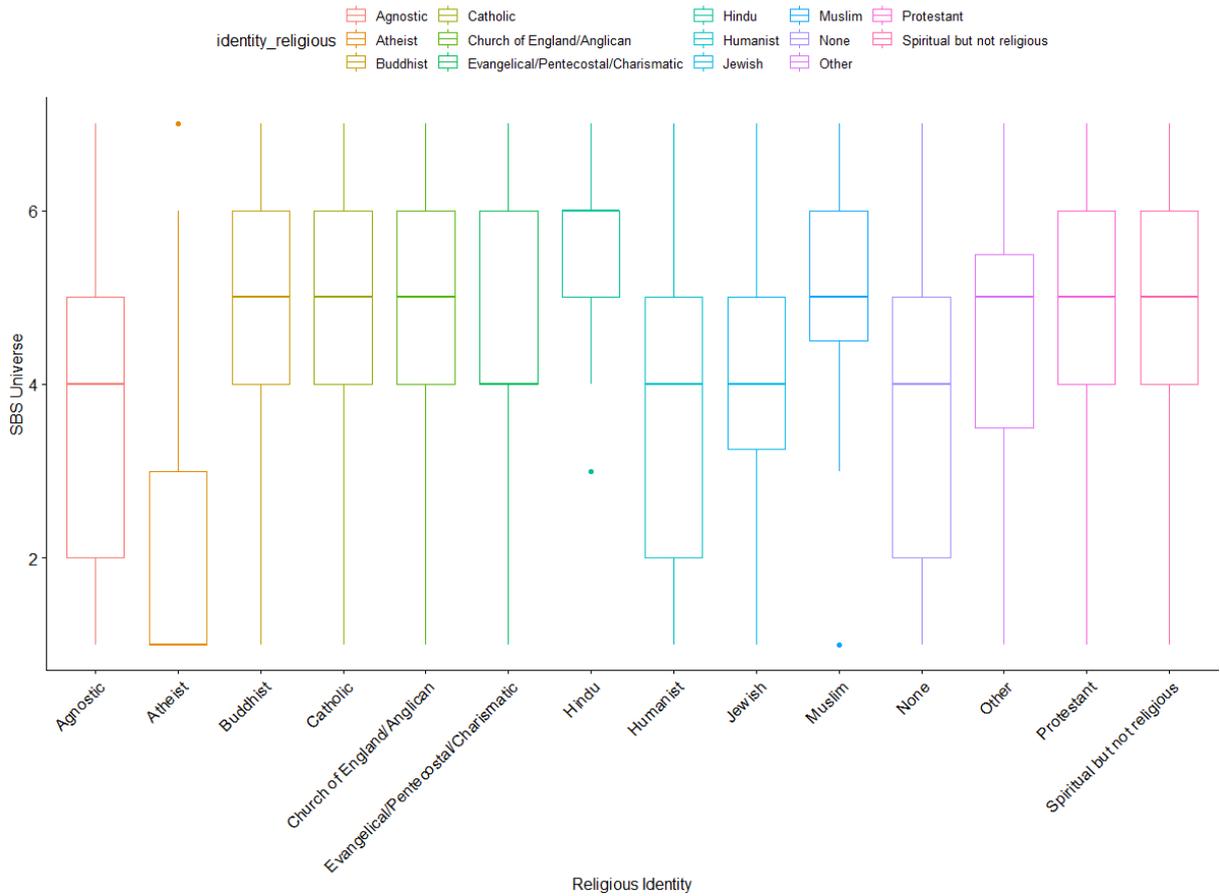

A similar pattern can be seen for the idea of karma, which appears to be more universally acceptable among all groups. However, we did find significant differences in the strength of support between religious groups ($F(13,2000) = 80.55$, $p < .01$)). This is particularly interesting because it is a hallmark of eastern religions but appears acceptable to some western nonreligious individuals. The distribution of belief in karma is presented in the figure below.

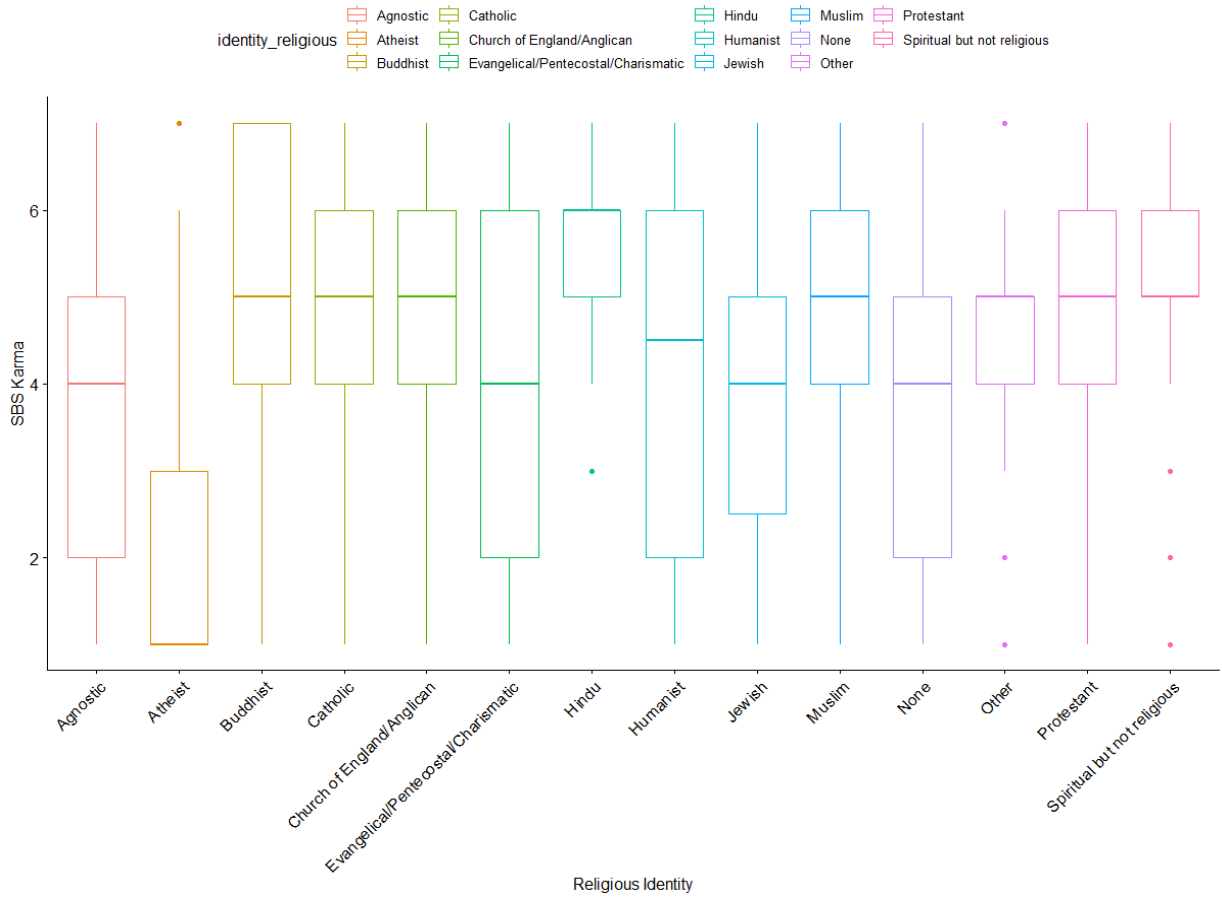

*Political Identity and Beliefs*

We found that our sample had a nearly uniform distribution of economic values, with there being a slight overrepresentation of people stating that they are "very liberal" in economic values, and a lower number of those stating that they are "very conservative".

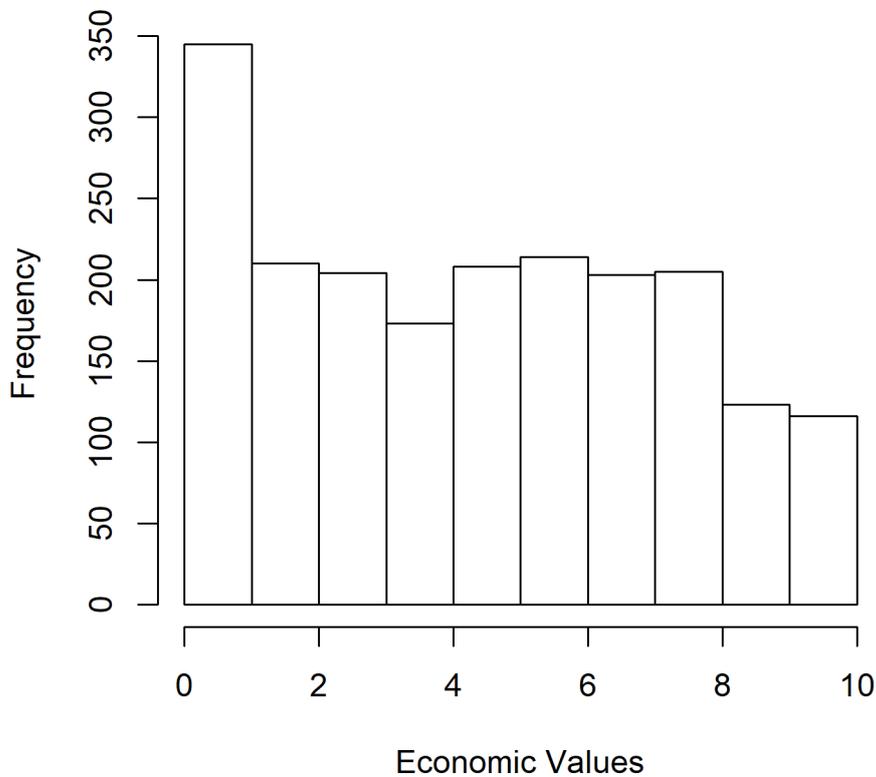

This over-representation of very liberal participants was exacerbated when it comes to the participants' self-reports on social values, where the distribution began to resemble a long tail distribution weighted toward very liberal social values. While this may not appear consistent with a general population, it is consistent with many online communities. To date, we are not aware of any investigation of the MTurk population in this regard; however, it is telling that it was overweighted in our sample. We do not believe that the recruitment of participants from social networks, which are well-known to lean more liberal in political persuasion, can explain this overall distribution.

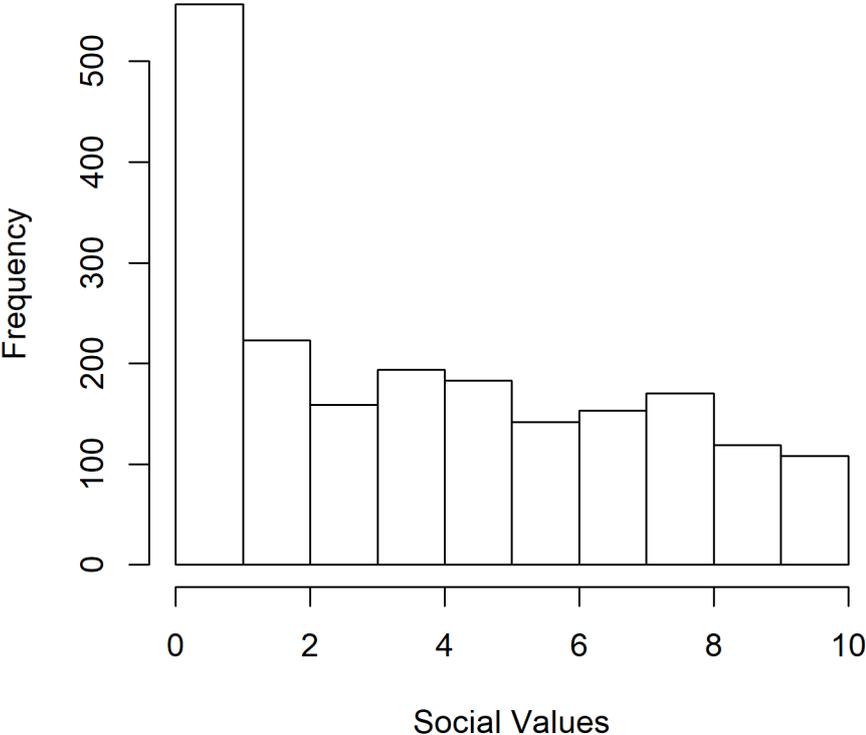

To further explore the political dimensions of the dataset we also mapped the social and economic dimensions along two axes to create a general heatmap of how these two variables correlate.

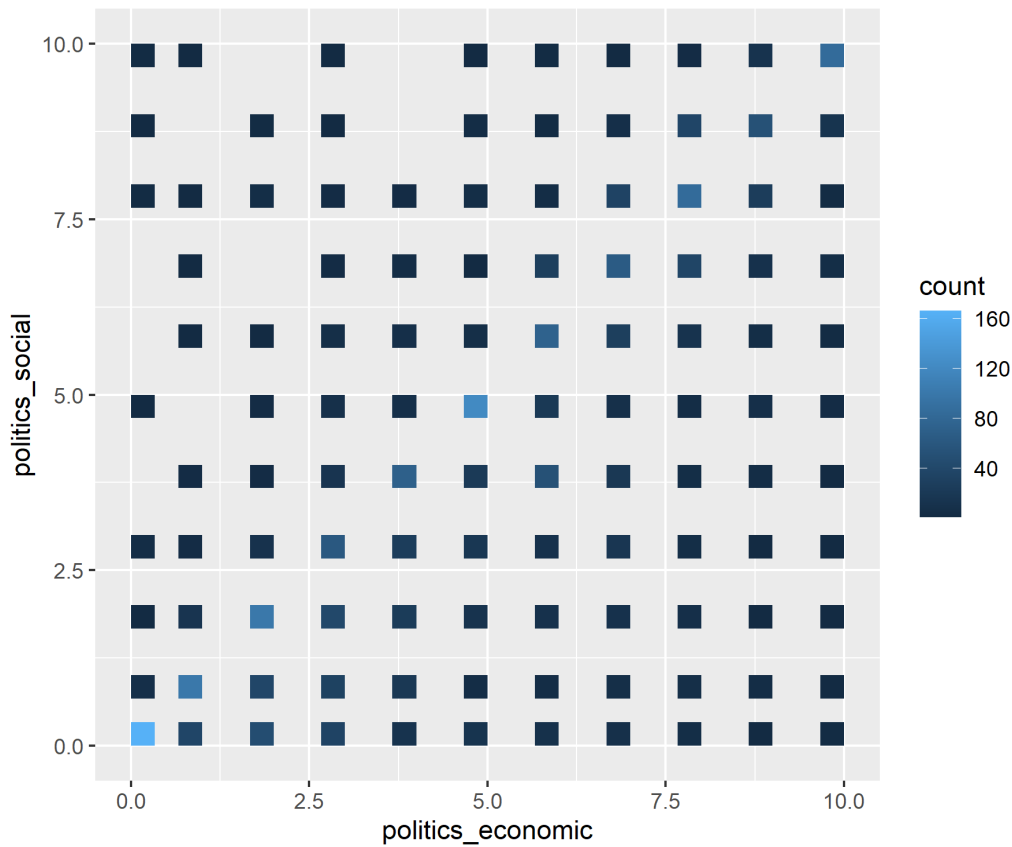

As expected, there is generally a strong correlation between the two dimensions of political ideologies. However, there is a sparser representation of individuals who say that they are very social conservative but economically liberal, whereas there is a well distributed segment of our sample that describe themselves as socially liberal but economically conservative. This is generally consistent with the political makeup of those in the online community, particularly from the UK and US, where a non-negligible, but small, group of people appear to adhere to ideologies that could be described as "classically liberal" or libertarian. This was further bolstered by the question of which political party they have supported; we found that there were many who supported third party candidates that aligned with those ideologies.

The table below shows frequency of support for all parties who reported support from more than 15 participants in our sample.

| Party | Frequency of support |
|---|---|
| Democrat | 639 |
| Republican | 316 |
| Labour | 71 |
| BIB | 64 |
| Conservative | 39 |
| Liberal | 35 |
| Green Party | 33 |
| Libertarian | 30 |
| Independent | 25 |
| Social Democrats | 24 |
| Lib Dem | 16 |

In assessing the responses of those who voted in the UK referendum we found the following:

| Position | Frequency |
|---|---|
| I voted leave and I still support my decision | 30 |
| I voted leave but regret my decision | 10 |
| I voted remain and I regret my decision | 3 |
| I voted remain and I still support my decision | 75 |

*Nationalism*

We captured a wide distribution of responses to the nationalism scale. While the distribution appears normal, it is also possible that because of the social unrest surrounding COVID-19 that we are seeing a muted distribution where national identity might be suppressed or re-evaluated, causing there to be more individuals on the extremes of the distribution than in the middle, effectively flattening our normal distribution.

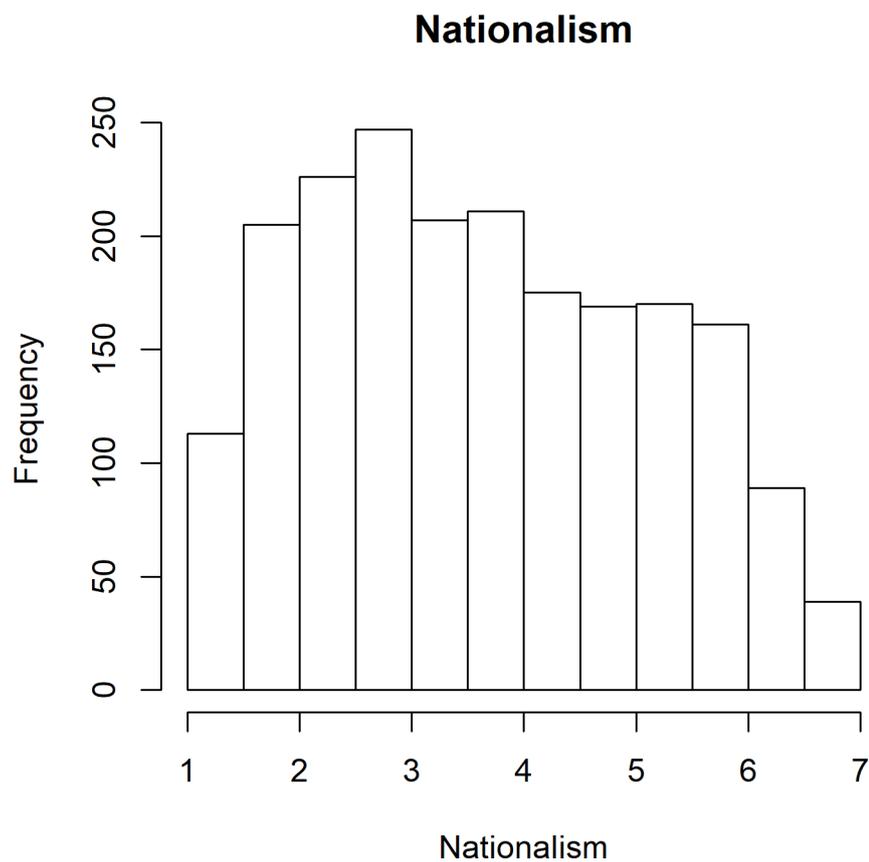

Generally, we found that there were significant differences between religious groups regarding nationalism. The tendency for some religious groups to be more nationalistic than others indicates that the nonreligious are also more likely to be non-nationalist; however, this varies by group, because humanists, for example, are not that different from Protestants generally.

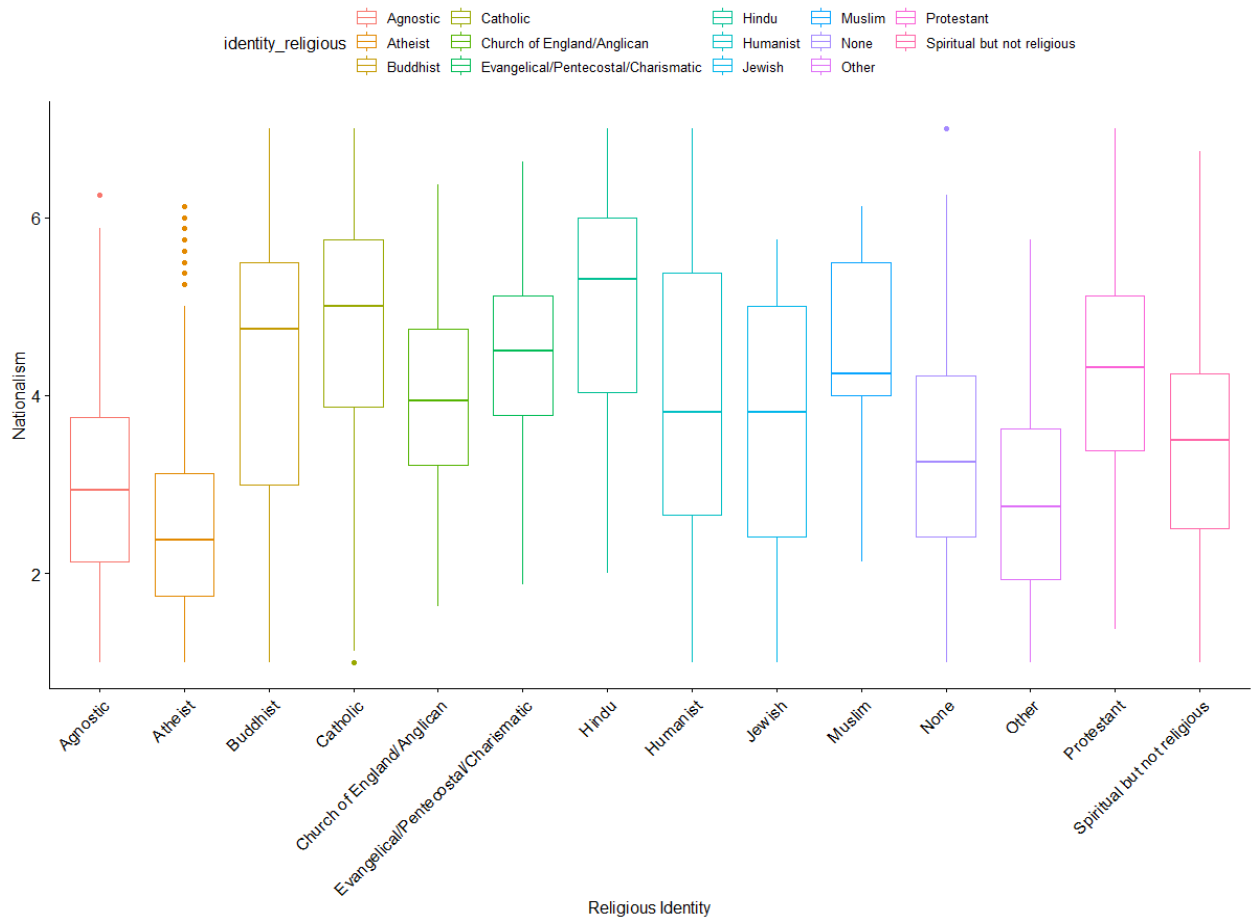

One pattern that we found interesting was the levels of nationalism by age. Graphing the participants' nationalism responses by age, we see that nationalism appears to peak among those in their mid 20s.

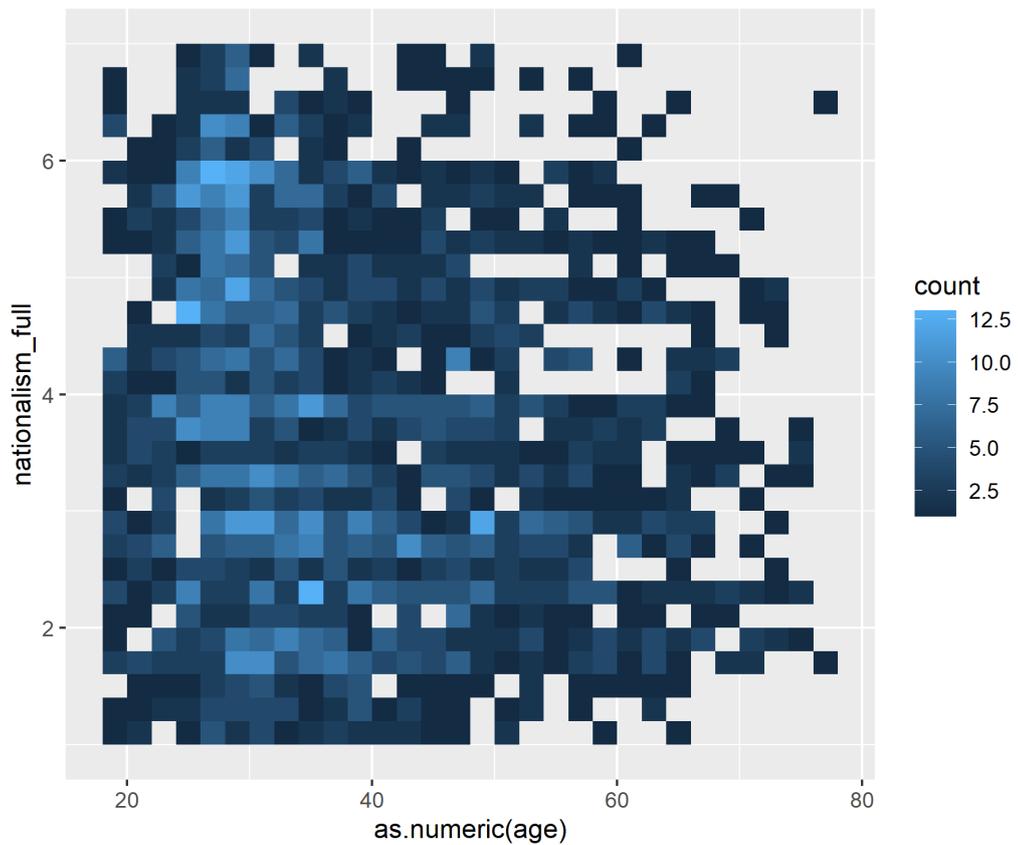

We further investigated the nationalism measure by looking at males and females separately. For males, we find that the hotspot tends to hold where the greatest nationalist cluster appears at around the age of 25. But nationalism continues well into middle age for men.

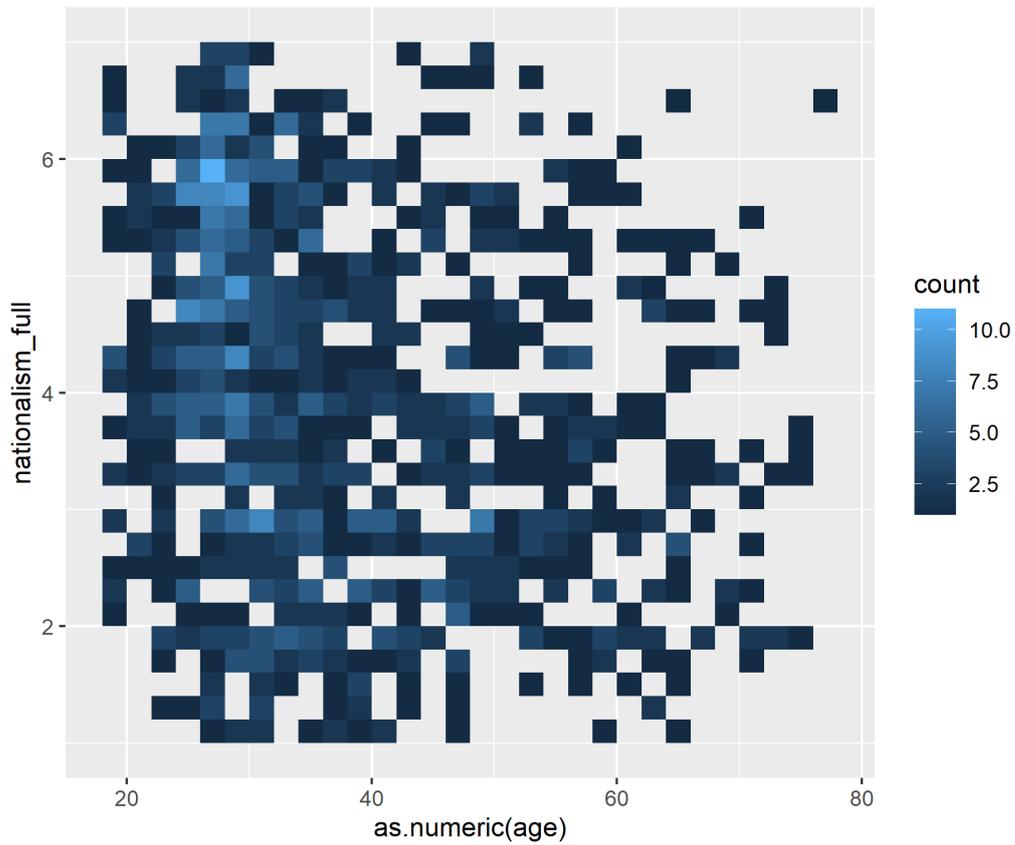

For females, we see a far less even distribution of nationalism; females tend to cluster lower in the nationalism scale, but the most nationalist females are found around the age of 30. Generally, female nationalism clusters in the lower levels of nationalism with the greatest number appearing below the midpoint; in opposition to the pattern exhibited by males.

Future research is needed to investigate why it is that nationalist tendencies appear different for each gender, and largely correspond to periods of life that are implicated in biological reproduction.

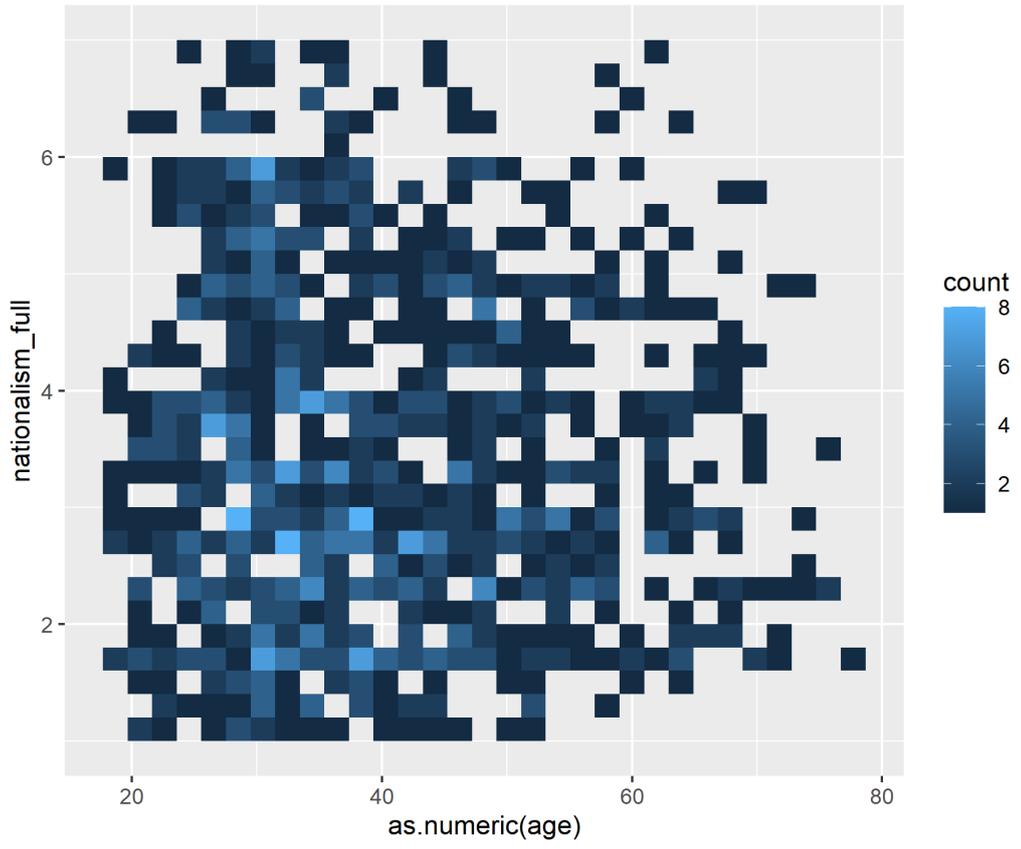

When assessing our threat measures, we found that there was a normal distribution of threat responses in our sample and that the additional financial threat measure did increase the overall threat perception.

When attempting to predict nationalism as a function of threat, we also found that the different threats had very different effects on nationalism, and that controlling for the threat that they believed COVID-19 posed to their way of life increased the model's predictability significantly (p<.01).

**Nationalism and Threat**

|  | Dependent variable: | |
|---|---|---|
|  | nationalism_full | |
|  | (1) | (2) |
| threat_predation | 0.197*** | 0.194*** |
|  | (0.034) | (0.033) |
| threat_contagion | -0.417*** | -0.398*** |
|  | (0.033) | (0.033) |
| threat_financial | -0.210*** | -0.219*** |
|  | (0.031) | (0.030) |
| threat_natural | -0.310*** | -0.379*** |
|  | (0.035) | (0.034) |
| threat_social | 0.572*** | 0.518*** |
|  | (0.040) | (0.039) |
| waylifethreatcovid |  | 0.290*** |
|  |  | (0.027) |
| Constant | 4.927*** | 4.200*** |
|  | (0.184) | (0.191) |
| Observations | 2,000 | 2,000 |
| $R^2$ | 0.245 | 0.287 |
| Adjusted $R^2$ | 0.244 | 0.285 |
| Residual Std. Error | 1.302 (df = 1994) | 1.266 (df = 1993) |
| F Statistic | 129.730*** (df = 5; 1994) | 133.875*** (df = 6; 1993) |

Note: *p<0.1; **p<0.05; ***p<0.01

**Threat**

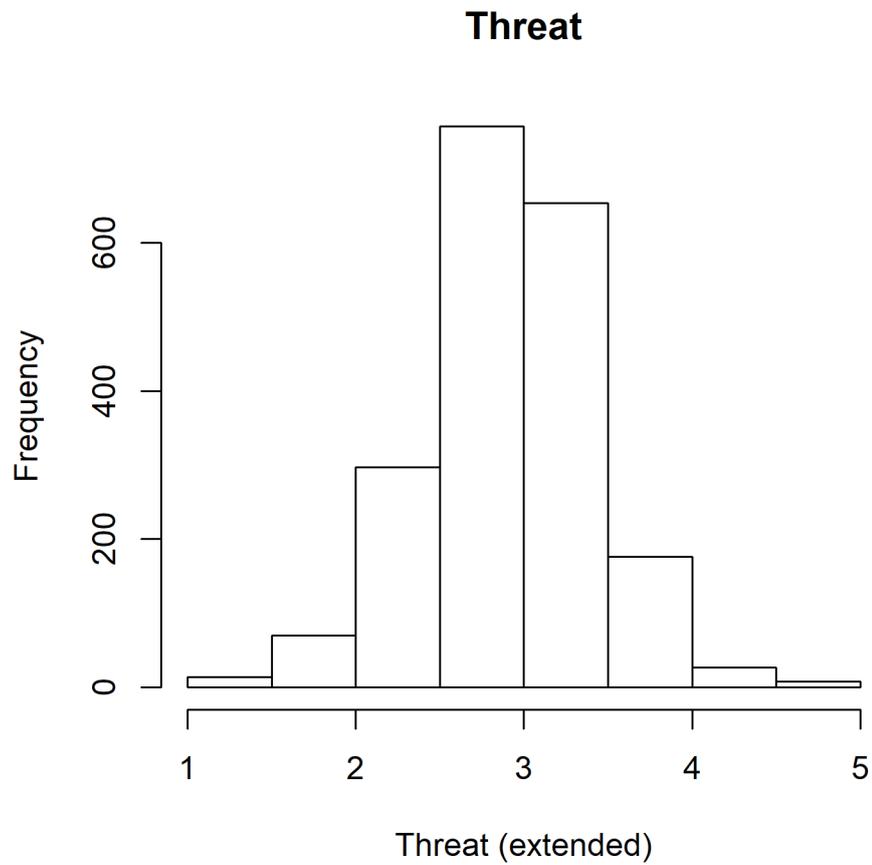

One of the things that stood out to us was that, as a scale, our threat measure did not have good validity as a single measure. Therefore, future analyses will need to investigate threat in relation to the sub-components of the measure; namely: contagion, predation, social, natural, and financial.

Our analysis uncovered a significant difference in perceived threat by gender.

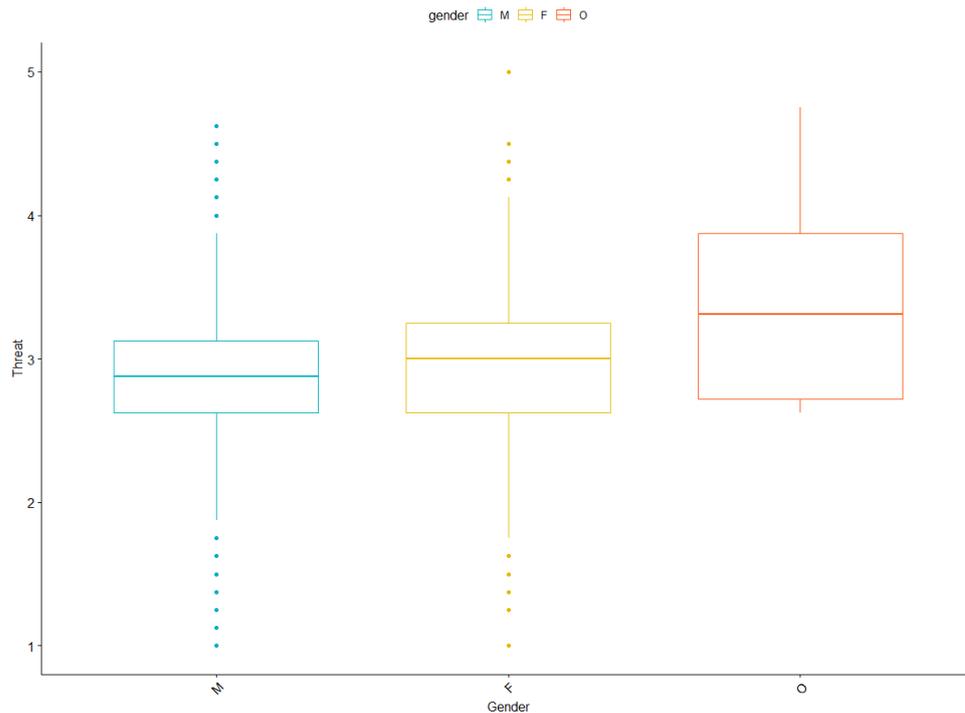

An ANOVA revealed that there was a significant difference between groups ($F(2,2002) = 10.25$, $p < .01$)). A posthoc test revealed that the effect was driven by all genders having significant differences (in all pairwise comparisons in a Tukey test $P<.01$).
When analysing the difference in threat perception by nationality, we did find that there was a significant difference between nationalities ($F(10,1994) = 1.84$, $p = .05$)) but a posthoc analysis revealed that the finding is driven exclusively by the difference between Spain and Canada. As such, we are inclined to reject the findings of that model as spurious and continue analyses as pooled given the small sample sizes drawn from those countries and their lesser relevance to the overall research question. We found no significant differences in threat perception by political party support ($p = .56$).

Given the interesting threat patterns discerned in earlier phases of the analyses, we explored the extent to which threat can explain other aspects of the survey results.
Using a General Linear Model (GLM) to assess the extent to which participant responses were affected by threat perceptions, we found that contagion and financial threats have a significant negative relationship with positive attitudes toward immigrants, while social threats have a positive relationship (reminder: the immigration question was binary 0 = no issue; 1 = would not want immigrant neighbours). This suggests that those who perceive more social threats are less likely to want immigrant neighbours, but those who perceive more contagion or financial threats are not.

| Immigrant Sentiment and Threats | |
|---|---|
| | Dependent variable: |
| | wvs_undes_immigrants |
| threat_predation | 0.040 |
| | (0.029) |
| threat_contagion | -0.087*** |
| | (0.029) |
| threat_financial | -0.072*** |
| | (0.027) |
| threat_natural | -0.038 |
| | (0.030) |
| threat_social | 0.113*** |
| | (0.034) |
| Constant | 0.620*** |
| | (0.159) |
| Observations | 2,002 |
| Log Likelihood | -3,084.607 |
| Akaike Inf. Crit. | 6,181.213 |
| Note: | *p<0.1; **p<0.05; ***p<0.01 |

In addition, we wanted to explore the effects that different threats had on supernatural beliefs. We found that threats have significantly different kinds of effects on different beliefs overall. Although all threats were found to be significant, predation and social threats were found to have a positive effect on supernatural beliefs, while contagion, financial, and natural threats were found to have a negative effect (results for both the original SBS and our extended SBS, where we added items for karma and universal force, are found below).

| Supernatural Beliefs and Threat | | |
|---|---|---|
| | Dependent variable: | |
| | sbs_orig_full | sbs_extended_full |
| | (1) | (2) |
| threat_predation | 0.243*** | 0.243*** |
| | (0.046) | (0.046) |
| threat_contagion | -0.638*** | -0.638*** |
| | (0.046) | (0.046) |
| threat_financial | -0.142*** | -0.142*** |
| | (0.042) | (0.042) |
| threat_natural | -0.299*** | -0.299*** |
| | (0.047) | (0.047) |
| threat_social | 0.488*** | 0.488*** |
| | (0.054) | (0.054) |
| Constant | 5.433*** | 5.433*** |
| | (0.252) | (0.252) |
| Observations | 2,002 | 2,002 |
| $R^2$ | 0.174 | 0.174 |
| Adjusted $R^2$ | 0.172 | 0.172 |
| Residual Std. Error (df = 1996) | 1.787 | 1.787 |
| F Statistic (df = 5; 1996) | 84.269*** | 84.269*** |
| Note: | *p<0.1; **p<0.05; ***p<0.01 | |

Lastly, we investigated the extent to which threats predicted the strength of different religious and national identification styles (namely, identification and fusion). We found an interesting pattern of results whereby for identification (with both religious and national affiliations) contagion and financial threats had a negative effect on identification, while social threats had a positive effect. Natural threats had a negative effect on national identification, but not religious identification; although it should be noted that this effect was extremely weak. All of the threats significantly predicted fusion, with predation and social threats having a positive effect on fusion with both religion and national groups, while contagion, financial, and natural threats had negative effects.

**Religious and National Identification and Threat**

|  | Dependent variable: | | | |
|---|---|---|---|---|
|  | sid_religious_full | ift_religious_full | sid_nation_full | ift_nation_full |
|  | (1) | (2) | (3) | (4) |
| threat_predation | 0.037 | 0.222*** | 0.010 | 0.168*** |
|  | (0.043) | (0.042) | (0.037) | (0.039) |
| threat_contagion | -0.076* | -0.359*** | -0.393*** | -0.505*** |
|  | (0.043) | (0.042) | (0.036) | (0.039) |
| threat_financial | -0.080** | -0.179*** | -0.151*** | -0.197*** |
|  | (0.039) | (0.038) | (0.033) | (0.036) |
| threat_natural | 0.040 | -0.149*** | -0.070* | -0.197*** |
|  | (0.044) | (0.043) | (0.037) | (0.040) |
| threat_social | 0.194*** | 0.438*** | 0.144*** | 0.347*** |
|  | (0.051) | (0.050) | (0.043) | (0.046) |
| Constant | 4.390*** | 4.414*** | 6.628*** | 5.923*** |
|  | (0.234) | (0.229) | (0.198) | (0.213) |
| Observations | 2,000 | 1,997 | 1,995 | 1,988 |
| $R^2$ | 0.011 | 0.110 | 0.082 | 0.149 |
| Adjusted $R^2$ | 0.008 | 0.108 | 0.080 | 0.147 |
| Residual Std. Error | 1.659 (df = 1994) | 1.624 (df = 1991) | 1.406 (df = 1989) | 1.503 (df = 1982) |
| F Statistic | 4.286*** (df = 5; 1994) | 49.365*** (df = 5; 1991) | 35.719*** (df = 5; 1989) | 69.380*** (df = 5; 1982) |
| Note: | | | | *p<0.1; **p<0.05; ***p<0.01 |

In addition to aspects of identification, we also investigated the role of threats in different personality traits. The regressions for each personality trait and each threat are presented in the table below.

**Big-5 Personality Variables and Threat**

| | Dependent variable: | | | | |
|---|---|---|---|---|---|
| | Openness_full | Conscientiousness_full | Extraversion_full | Agreeableness_full | Neuroticism_full |
| | (1) | (2) | (3) | (4) | (5) |
| threat_predation | -0.026 | -0.064*** | 0.003 | -0.071*** | -0.151*** |
| | (0.022) | (0.019) | (0.026) | (0.022) | (0.026) |
| threat_contagion | 0.059*** | -0.087*** | -0.037 | -0.171*** | -0.163*** |
| | (0.022) | (0.019) | (0.026) | (0.022) | (0.026) |
| threat_financial | 0.064*** | -0.044** | -0.087*** | -0.114*** | -0.098*** |
| | (0.020) | (0.017) | (0.024) | (0.020) | (0.024) |
| threat_natural | 0.145*** | 0.081*** | -0.017 | 0.051** | 0.126*** |
| | (0.023) | (0.019) | (0.027) | (0.023) | (0.027) |
| threat_social | -0.197*** | -0.098*** | -0.027 | -0.167*** | -0.226*** |
| | (0.026) | (0.022) | (0.031) | (0.026) | (0.031) |
| Constant | 3.177*** | 4.459*** | 3.425*** | 4.601*** | 4.342*** |
| | (0.120) | (0.102) | (0.141) | (0.121) | (0.144) |
| Observations | 1,999 | 2,000 | 2,001 | 2,000 | 2,000 |
| $R^2$ | 0.071 | 0.058 | 0.012 | 0.110 | 0.123 |
| Adjusted $R^2$ | 0.068 | 0.056 | 0.010 | 0.108 | 0.121 |
| Residual Std. Error | 0.850 (df = 1993) | 0.726 (df = 1994) | 1.002 (df = 1995) | 0.857 (df = 1994) | 1.018 (df = 1994) |
| F Statistic | 30.319*** (df = 5; 1993) | 24.528*** (df = 5; 1994) | 4.902*** (df = 5; 1995) | 49.488*** (df = 5; 1994) | 56.031*** (df = 5; 1994) |

Note: *p<0.1; **p<0.05; ***p<0.01

No further analysis was undertaken because all of the relationships are relatively weak. Explanatory power is therefore low. Moreover, there are debates as to what causal directions, or even feedback loops, might be proposed theoretically. It is very clear that further research is needed.

*Structural Equation Model*

To further asses the data and begin to find structure in the data that could be usable for informing our system dynamics model, we utilized structural equation modelling to test several models that we deemed as theoretically interesting or plausible. Variations on models were found as an interesting pattern about the threat systems. We found in our analysis that threat does not operate well as a single measure—as suggested above. However, we found that threat does function acceptably well as a multi-dimensional scale, with two clusters. In our structural equation modelling, we found that two clusters of threats have two different and opposing effects on social and economic conservatism. The model that we used to inform a great deal of our system dynamics model is depicted below.

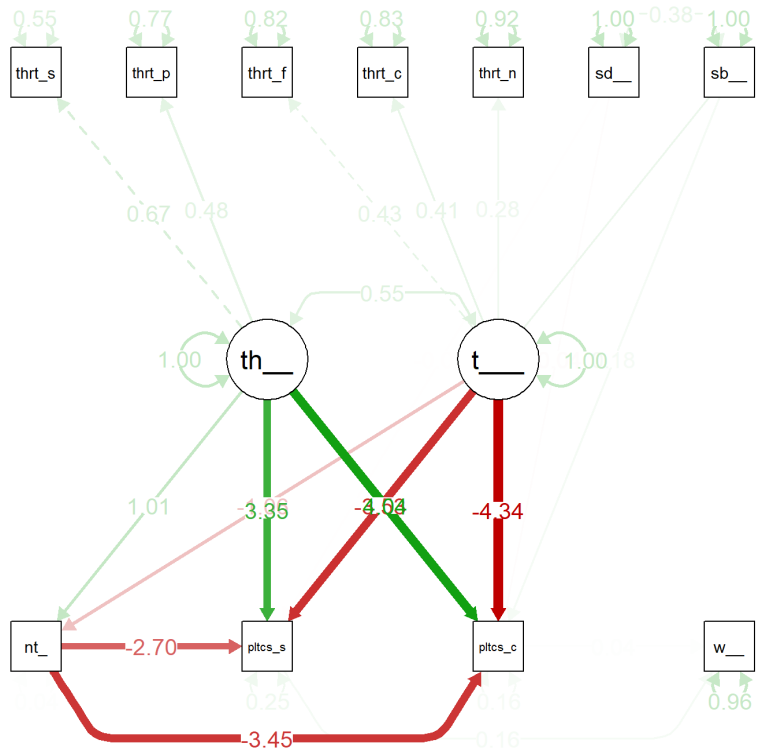

Along the top we have the 5 threats as well as supernatural beliefs and identity. The two large circles in the middle are latent variables constructed to capture the co-variance in the sub-items of the threat measure. Along the bottom we have nationalism, social and economic conservativism. Lastly, on the lower right we have the WVS measure for anti-immigrant values. As indicated in the figure, social and predation threats form one cluster, which has a positive effect on social and economic conservativism. Meanwhile, financial, contagion and natural threats have a negative effect, predicting higher levels of liberal political beliefs. These two effects had similar directional effects on nationalism, which also had a negative effect on social conservativism, but a positive effect on anti-immigrant sentiment.

Statistical analysis of the model above found that it has an acceptable fit (statistical analyses provided in the table below). This model performed better than most of the other models that we tested (and had an AIC of 55566.78); therefore, we felt that it was the most generalizable theoretical model of those investigated for this pilot study.

|  | Model Estimate | Std. Err. | z | p |
|---|---|---|---|---|
| | | Factor Loadings | | |
| threat_soc_pred | | | | |
| threat.social | 1.00[+] | | | |
| threat.predation | 0.81 | 0.07 | 12.36 | .000 |
| threat_cont_fin_nat | | | | |
| threat.financial | 1.00[+] | | | |
| threat.contagion | 0.86 | 0.11 | 7.47 | .000 |
| threat.natural | 0.56 | 0.11 | 5.36 | .000 |
| | | Regression Slopes | | |
| nationalism_full | | | | |
| threat.soc.pred | 2.75 | 0.25 | 11.20 | .000 |
| threat.cont.fin.nat | -3.62 | 0.26 | -13.70 | .000 |
| politics_social | | | | |
| threat.soc.pred | 17.52 | 32.95 | 0.53 | .595 |
| threat.cont.fin.nat | -23.23 | 38.15 | -0.61 | .543 |
| nationalism.full | -5.18 | 10.58 | -0.49 | .624 |
| sid.religious.full | -0.04 | 0.03 | -1.21 | .226 |
| sbs.extended.full | 0.48 | 0.04 | 13.70 | .000 |
| politics_economic | | | | |
| threat.soc.pred | 20.82 | 28.14 | 0.74 | .459 |
| threat.cont.fin.nat | -28.20 | 37.85 | -0.75 | .456 |
| nationalism.full | -6.51 | 9.47 | -0.69 | .492 |
| sid.religious.full | -0.08 | 0.04 | -2.13 | .033 |
| sbs.extended.full | 0.26 | 0.04 | 7.00 | .000 |
| wvs_undes_immigrants | | | | |
| politics.social | 0.06 | 0.01 | 5.01 | .000 |
| politics.economic | 0.02 | 0.01 | 1.15 | .249 |
| | | Residual Variances | | |
| threat.social | 0.38 | 0.03 | 13.74 | .000 |
| threat.predation | 0.67 | 0.03 | 24.36 | .000 |
| threat.financial | 0.87 | 0.04 | 23.42 | .000 |
| threat.contagion | 0.68 | 0.03 | 25.96 | .000 |
| threat.natural | 0.71 | 0.03 | 28.06 | .000 |
| nationalism.full | 0.08 | 0.04 | 1.87 | .061 |
| politics.social | 2.06 | 1.98 | 1.04 | .299 |
| politics.economic | 1.29 | 3.01 | 0.43 | .668 |
| wvs.undes.immigrants | 1.24 | 0.08 | 15.90 | .000 |
| sid.religious.full | 2.76[+] | | | |
| sbs.extended.full | 3.87[+] | | | |

|  | Residual Covariances |  |  |  |
|---|---|---|---|---|
| sid.religious.full w/sbs.extended.full | 1.23[+] |  |  |  |
|  | Latent Variances |  |  |  |
| threat.soc.pred | 0.30 | 0.03 | 10.43 | .000 |
| threat.cont.fin.nat | 0.19 | 0.04 | 5.27 | .000 |
|  | Latent Covariances |  |  |  |
| threat.soc.pred w/threat.cont.fin.nat | 0.13 | 0.03 | 4.59 | .000 |
|  | Fit Indices |  |  |  |
| $\chi^2$ | 1588.64(34) |  |  | .000 |
| CFI | 0.74 |  |  |  |
| TLI | 0.58 |  |  |  |
| RMSEA | 0.15 |  |  |  |

[+]Fixed parameter

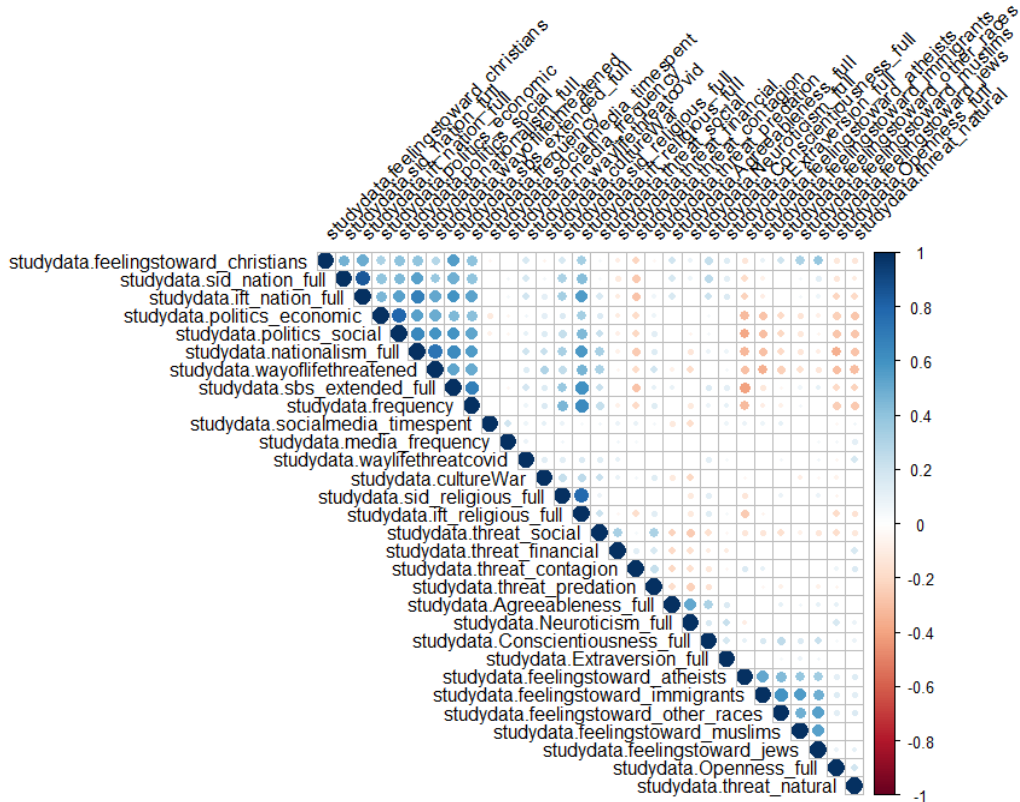

In addition to the SEM, we also provide a full correlation heat map of all of the independent correlations found in our dataset (see above).

Lastly, we utilized a machine learning system to highlight what features could help us classify participants based on their responses to the WVS immigrant question. The results were as follows:

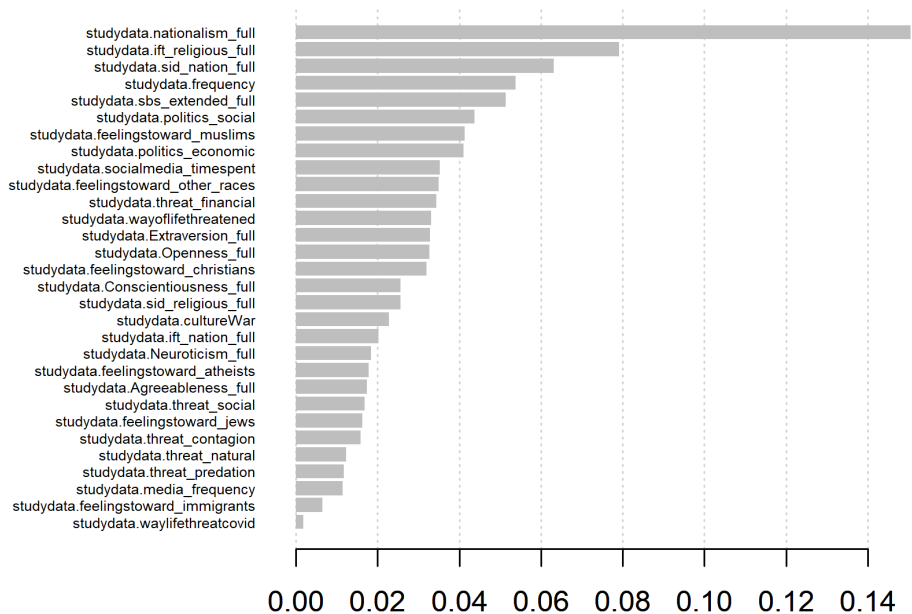

**Study 2**

Study 2 utilized the results of Study 1 to create a system dynamics model that aims to add causal propositions and links to the variables in a way that allows us to capture complex feedback dynamics and changes over time, as well as the way in which the system can be affected by psychologically realistic mechanisms, such as habituation to the threats that were so important in the earlier model.

**Methods**

To construct this system dynamics model, we utilized the software platform AnyLogic 8 (The AnyLogic Company, 2017). The model was written in that platform and functions were added in raw Java in order to allow for more accurate calculations and testing.

The simulation effectively has two subsystems that work in concert throughout the run of any given simulation.

The first system is designed to capture threat perception dynamics. The general form of the model is similar to the one presented and discussed in Shults, Lane, et al. (2018). It utilizes a system dynamics model that allows for habituation dynamics to be instantiated that can mimic the habituation system described by Rescorla & Wagoner. This threat perception system has 5 subsystems, one for each dimension of threat measured in this study. It is visually depicted in the figure below.

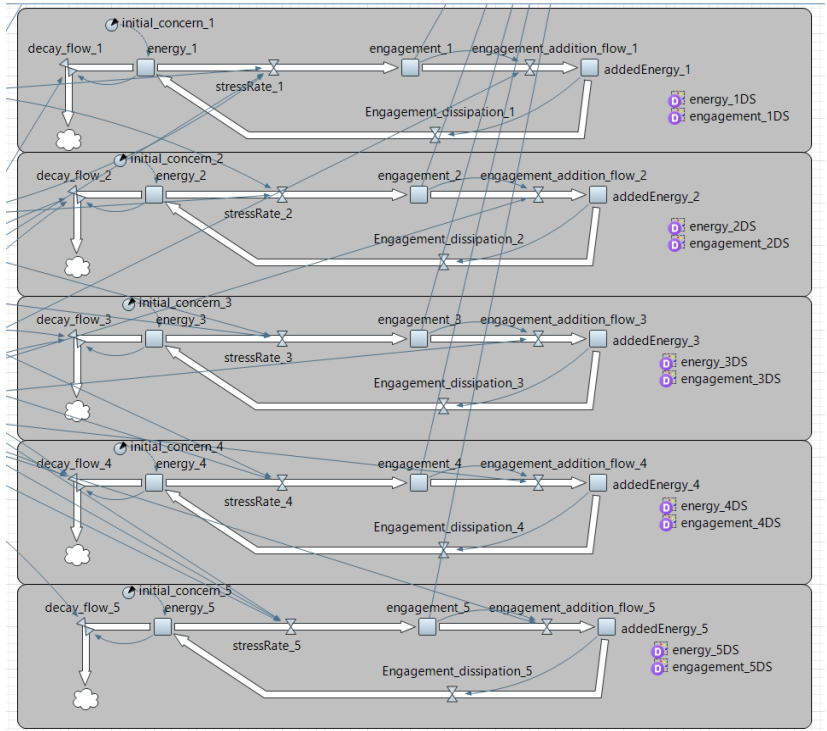

As the different levels of engagement with threats are calculated, their respective levels are then aggregated into two dynamic variables, mimicking the latent variables in the structural equation model presented in Study 1. These were then posited to interact with the different socio-political variables in the model and psychological tendencies and other individual level variables and patterns listed in the parameters in the model description below. The structure of the model is depicted in the figure below.

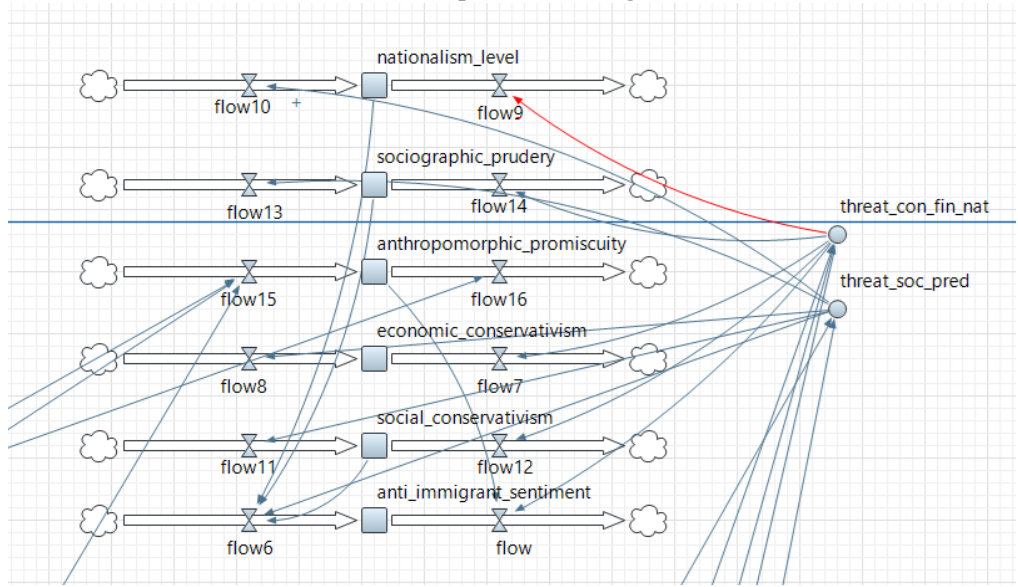

Most of the variables in the figure are self-explanatory, but two require brief clarification (for a more detailed description, see. As in the article on modelling terror management theory mentioned above (Shults, Lane, et al., 2018), we are operationalizing "religiosity" in this context to designate "socially shared cognitive and ritual engagement with axiologically relevant supernatural agents postulated within one's in-group." This sort of imaginative engagement, which promotes cooperation, commitment, and cohesion in the face of out-group threats and environmental challenges, is fostered by two reciprocally reinforcing evolved dispositions: the tendency to infer human-like supernatural causes and the tendency to prefer coalition-favouring moral prescriptions when confronted with ambiguous or frightening phenomena. In other words, religiosity involves the intensification and integration of a hyper-active propensity toward detecting gods as hidden agents and a hyper-active propensity toward protecting in-group norms. We refer to these as "anthropomorphic promiscuity" and "sociographic prudery" respectively (for theoretical background, see Shults, 2014, 2018).

A list of the stocks, parameters, and dynamic variables are presented in the table below. Full source code for the model can be found on GitHub[1] (currently a private repository until publication).

| Variable Type | Variable Name | Description |
| --- | --- | --- |
| Parameter | Big_5_agreeableness | |
| | Big_5_conscientiousness | |
| | Big_5_extraversion | |
| | Big_5_neuroticism | |
| | Big_5_openness | |
| | energyDecay | |
| | habituationRate | |
| | Hazard_intensity_contagion | |
| | Hazard_intensity_financial | |
| | Hazard_intensity_natural | |
| | Hazard_intensity_predation | |
| | Hazard_intensity_social | |
| | Initial_concern_1 | |
| | Initial_concern_2 | |
| | Initial_concern_3 | |
| | Initial_concern_4 | |
| | Initial_concern_5 | |
| | LHS_RUN | |
| | Rel_frequency | |
| | socialMediaUse | |
| | ThreatPctOfMedia | |
| | tvMediaUse | |

**Design of Experiment**

To test our model, we utilized a parameter sweep of the theoretical space of the model. This was achieved by selecting 20,000 uniformly distributed sets of variables for all of the parameters listed in the table above. This allowed us to statistically test the model for all theoretically relevant settings to better understand the causal links proposed in the model and explore the relations among the variables.

Each parameter set was input into the model when a simulation run started and was run once. Because of the non-stochastic nature of the system dynamics model we specified each parameter set only needed to be run once.

At the end of each run, data on the input parameters, as well as for each of the following outputs was saved for analysis: hazard_event_count_contagion, hazard_event_count_financial, hazard_event_count_natural, hazard_event_count_predation, hazard_event_count_social, nationalism_level,

---

[1] https://github.com/cogijl/kingstonThreatStudy

economic_conservativism, social_conservativism, anthropomorphic_promiscuity, sociographic_prudery, anti_immigrant_sentiment, threat_con_fin_nat, threat_soc_pred, engagement_1, engagement_2, engagement_3, engagement_4, engagement_5, energy_1, energy_2, energy_3, energy_4, energy_5, addedEnergy_1, addedEnergy_2, addedEnergy_3, addedEnergy_4, addedEnergy_5,

**Results**

The data output by the experimental parameter sweep described above was analysed in order to better understand the theoretical space of what "could be" using a combination of correlations, regressions, and visualizations. Not all the correlations from the survey are expected to be present in the simulated data because the latter is the result of a specifically defined complex computational model that is reflective of, but not matched by, the complexity of the real world from which the survey data was drawn.

The first analysis that was run was a correlation plot between all of the variables, similar to what was presented for the survey data.

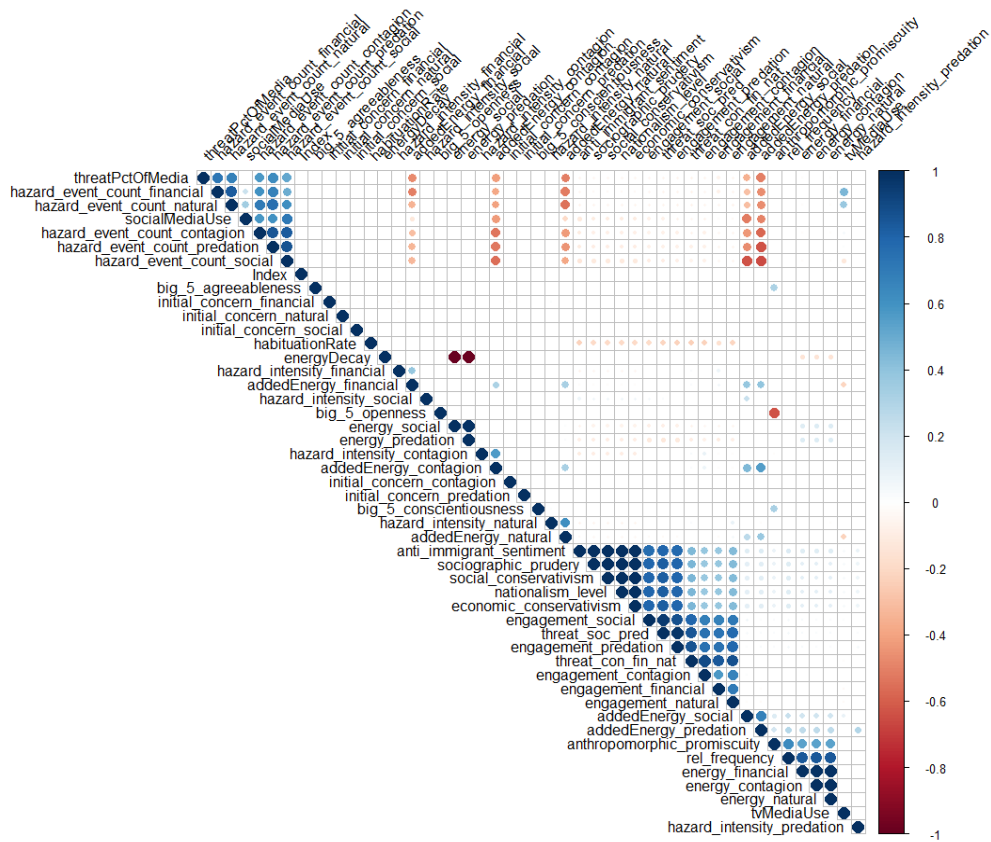

One of the ways that the simulation data can be investigated is through visual representations of relationships under certain partitions. For example, when visualizing immigrant sentiment (x axis) and social conservativism (y axis), but only for simulations where the nationalism was within the lowest quartile, we find that there are instances where extreme anti-immigrant sentiment is possible, but only among individuals who are socially hyper-liberal. This is a fascinating finding insofar as it challenges common sense, and the survey data to an extent, but it is consistent with the rise of xenophobia found historically in hyper-left Marxist states, for example. However, in the simulated data, as in history, that is generally rare.

As a point of validation, we wanted to see a clustering of causal effects reflected in our simulated data similar to those found in the survey data. Using a regression to test the effects of different threats on anti-immigrant sentiment, we did see the sort of clustering that we expected.

**Simulated threat and anti-immigrant relationships**

|  | Dependent variable: |
|---|---|
|  | anti_immigrant_sentiment |
| engagement_social | 694.885*** |
|  | (8.181) |
| engagement_financial | -1,047.842*** |
|  | (11.688) |
| engagement_contagion | -798.107*** |
|  | (8.454) |
| engagement_natural | -876.415*** |
|  | (15.046) |
| engagement_predation | 1,107.319*** |
|  | (6.956) |
| Constant | 1,864.321*** |
|  | (63.985) |
| Observations | 20,000 |
| $R^2$ | 0.846 |
| Adjusted $R^2$ | 0.846 |
| Residual Std. Error | 8,891.673 (df = 19994) |
| F Statistic | 21,925.840*** (df = 5; 19994) |

Note: *p<0.1; **p<0.05; ***p<0.01

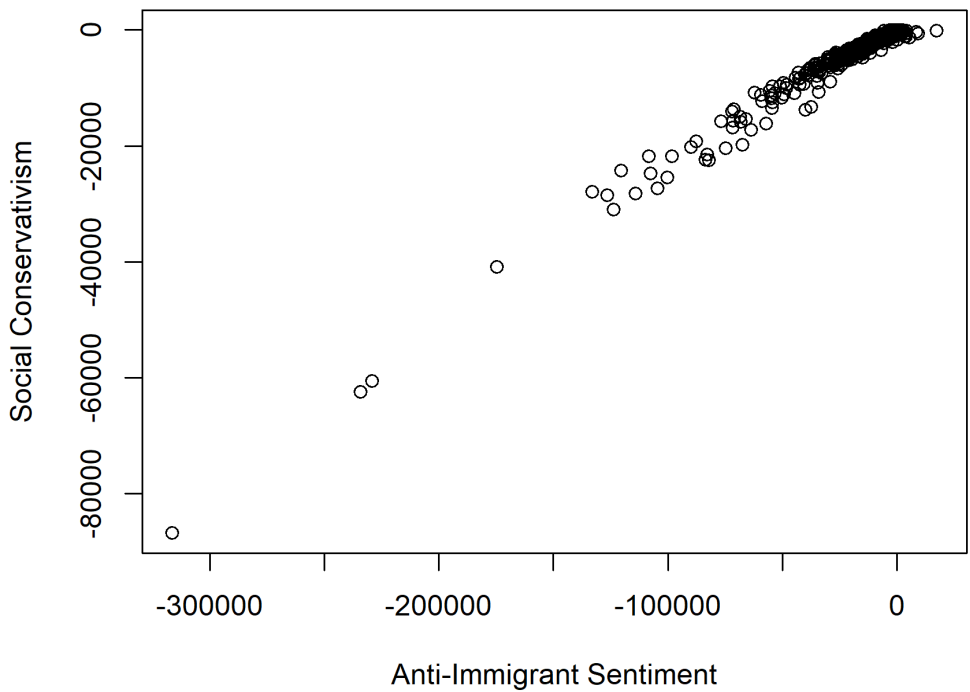

**Anti-immigrant sentiment X conservativism (non-nationalists)**

In addition, further extremes can be found in the relationship between anti-immigrant sentiment and media use. Among those who are the least frequent consumers of social media, it seems that social conservativism can be extremely high, as can anti-immigrant sentiment.

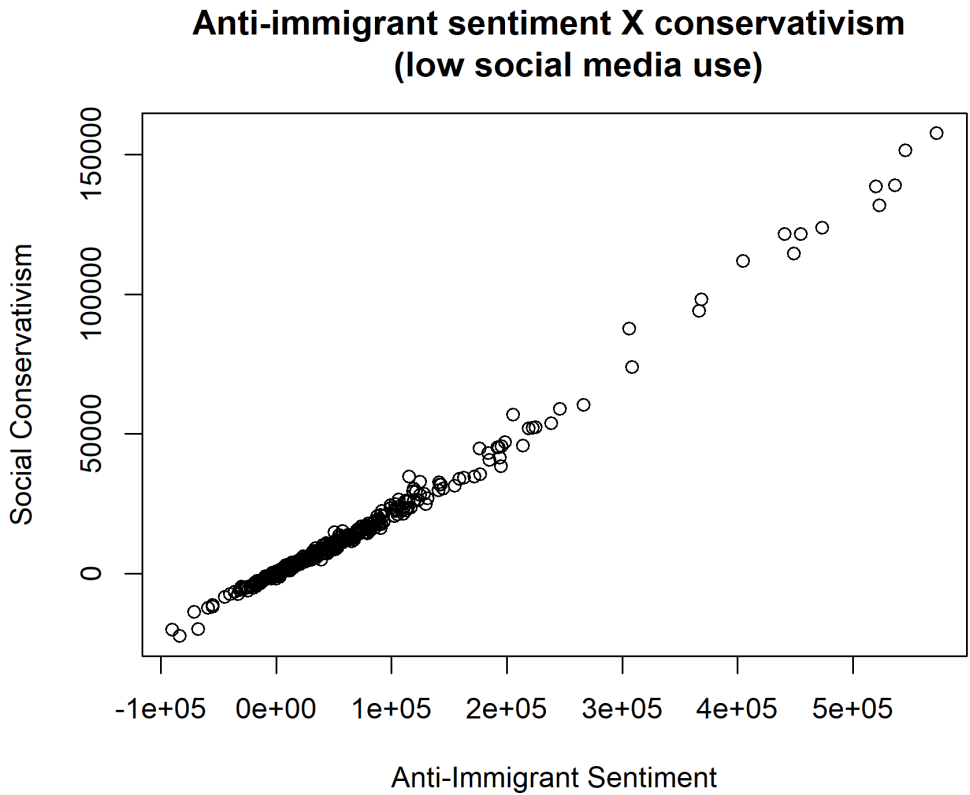

**Anti-immigrant sentiment X conservativism (low social media use)**

However, among the superusers of social media in the simulated data, we see that anti-immigrant sentiment can be both positive and negative, depending on the parameters of the simulation. The observed levels of social liberalism (the opposite of social conservativism in the model) are far more extreme than among those who use little social media. This was an unintended outcome but aligns well with earlier surveys and literature suggesting an over-representation of socially liberal ideological frameworks present in online social networks such as Twitter.

**Anti-immigrant sentiment X conservativism
(high social media use)**

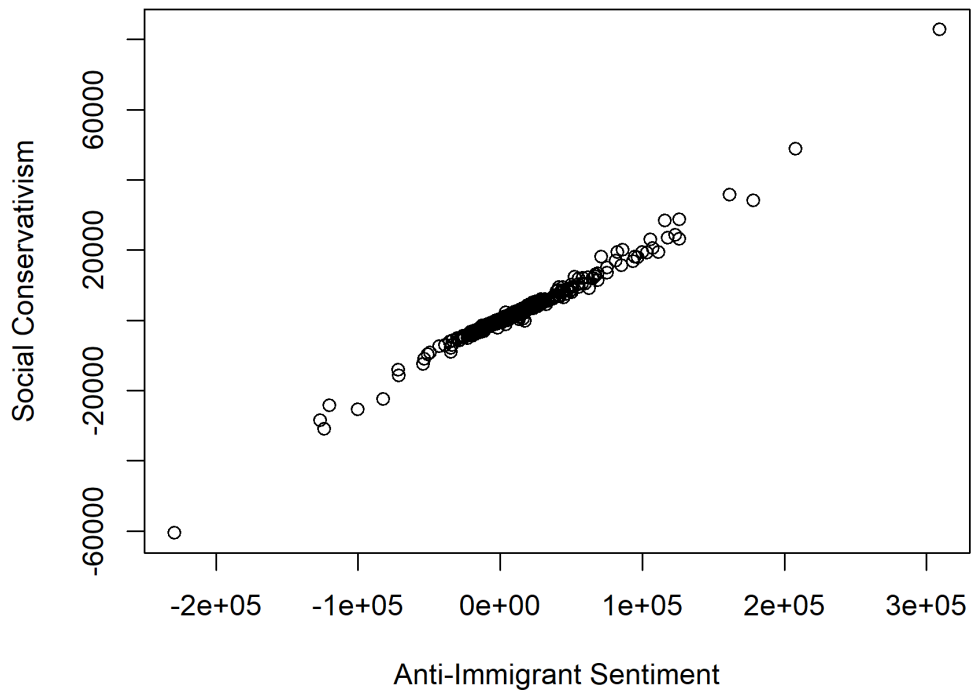

In our simulated data, we found a normal distribution of anthropomorphic promiscuity, with a central tendency toward a relatively low, but positive number. This suggests that given the total theoretical space, most simulations resulted in some god belief, but some simulations exhibiting extreme anti-or-pro god beliefs. This seems to reflect real world patterns as well.

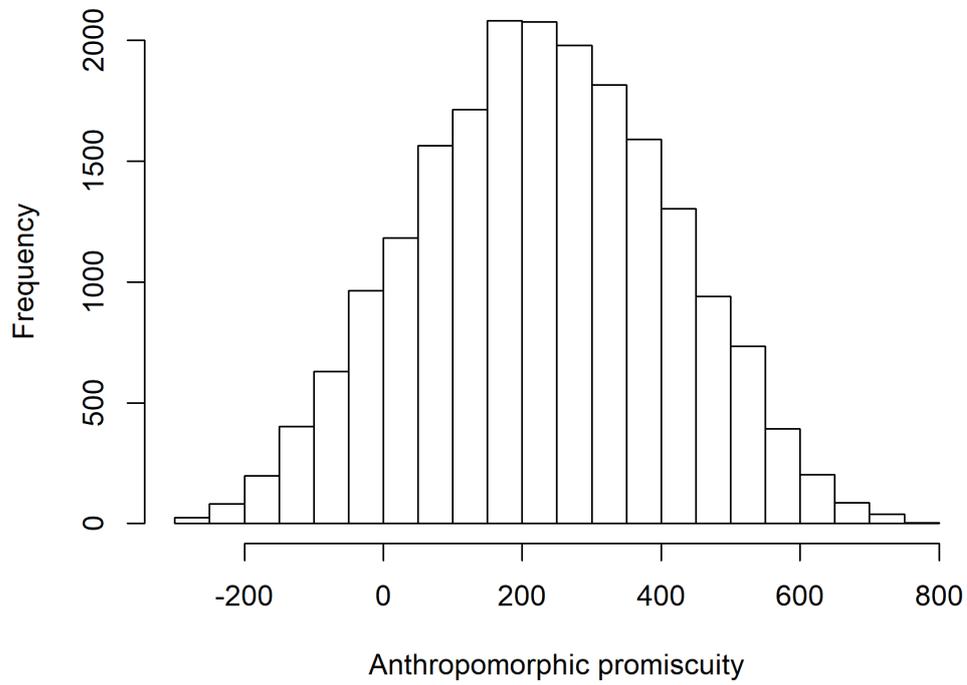

## Anthropomorphic promiscuity

Generally, engagement with different threats represented a long tail distribution, where most people engaged very rarely with threats; however, on rare occasions individuals engaged with extreme numbers of threats. An example from contagion threats is depicted below.

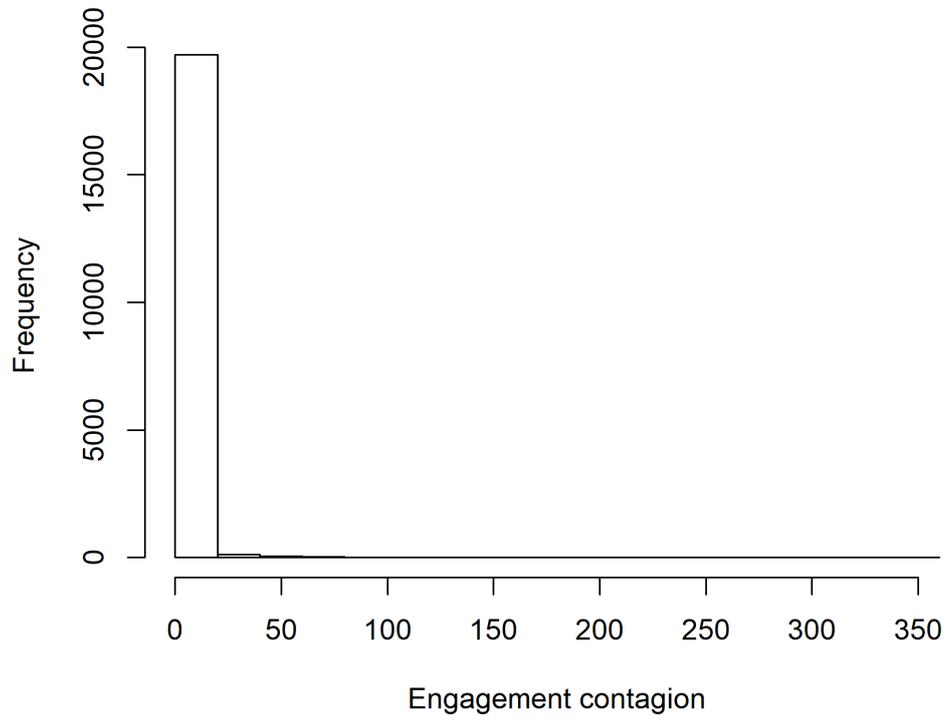

Since previous research has found nationalism to be a key indicator of anti-immigrant sentiment, we investigated levels of nationalism in the simulated data. Due to the correlation between nationalism and religious attendance in the survey data, we investigated this relationship in our simulated data (visualized below).

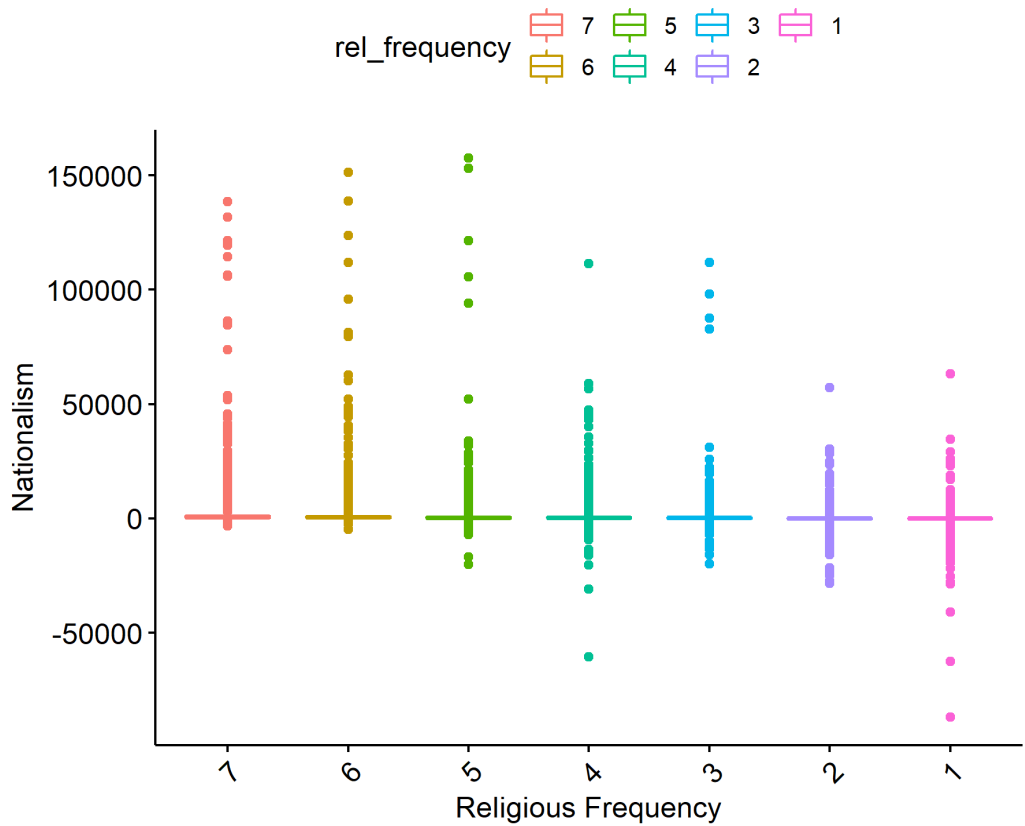

Generally, there is a clear trend where more frequent religious attendance can result in higher levels of nationalism. However, in all cases, the levels of nationalism tend to be relatively low. This suggests that, in our simulation, even though there is a positive relationship between religious attendance and nationalism, that it is not a key driver of nationalism, but has a relatively small effect on the system.

Additional regressions revealed several interesting patterns by sub-setting the data as being either high or low in nationalism. One of the results is that when nationalism is low, there is a significant effect on financial threats on anti-immigrant sentiment, but when nationalism is high there is no significant effect. This suggests that there is need for more research on the effect of the specific beliefs of a group on its members' acceptance of immigrants, and on the way in which individuals' framing of their national identity informs the extent to which they'll accept immigrants.

Results of these regressions can be found in the table below. In the table, model 1 is the entire sample of simulated data, model 2 is only those simulations resulting in low nationalism, and model 3 is only those simulations resulting in high nationalism.

| | Dependent variable: anti_immigrant_sentiment | | |
|---|---|---|---|
| | (1) | (2) | (3) |
| tvMediaUse | 155.293 | 199.861* | 311.240 |
| | (102.377) | (121.233) | (336.524) |
| threatPctOfMedia | -107.625 | -902.908*** | 462.309 |
| | (143.656) | (186.989) | (461.925) |
| socialMediaUse | -138.927 | -748.730*** | -52.424 |
| | (130.348) | (161.135) | (401.519) |
| rel_frequency | 273.552*** | 291.233*** | 766.765*** |
| | (9.637) | (17.281) | (40.094) |
| initial_concern_social | -45.323 | 69.218 | -186.211 |
| | (66.176) | (77.618) | (215.516) |
| initial_concern_financial | 14.363 | -91.998 | 194.417 |
| | (66.217) | (77.476) | (213.138) |
| initial_concern_contagion | 176.238*** | 246.201*** | 364.315* |
| | (66.192) | (77.158) | (214.463) |
| initial_concern_predation | -15.764 | -18.941 | -15.237 |
| | (66.195) | (77.017) | (212.145) |
| hazard_intensity_contagion | -1,532.591*** | -1,859.861*** | -2,506.320*** |
| | (66.518) | (103.222) | (239.938) |
| hazard_intensity_financial | -1,407.882*** | -1,515.354*** | -2,891.300*** |
| | (66.319) | (84.968) | (222.588) |
| hazard_intensity_natural | -885.580*** | -930.353*** | -1,638.606*** |
| | (66.296) | (88.660) | (219.454) |
| hazard_intensity_predation | 1,593.597*** | 1,668.852*** | 3,235.817*** |
| | (66.397) | (97.268) | (237.116) |
| hazard_intensity_social | 1,162.412*** | 1,046.521*** | 2,969.025*** |
| | (66.379) | (82.051) | (220.674) |
| habituationRate | -3,121.543*** | 1,626.501*** | -12,915.230*** |
| | (67.521) | (81.172) | (331.911) |
| energyDecay | 75.349 | 12.673 | 157.790 |
| | (66.211) | (76.883) | (211.905) |
| big_5_openness | 784.327*** | 820.899*** | 690.610*** |
| | (66.201) | (78.042) | (213.052) |
| big_5_conscientiousness | -408.887*** | -288.639*** | -668.946*** |
| | (66.215) | (78.188) | (210.549) |
| big_5_agreeableness | -322.506*** | -341.067*** | -150.314 |
| | (66.184) | (77.305) | (212.707) |
| hazard_event_count_contagion | -54.260*** | -12.868 | -204.400*** |
| | (11.212) | (12.095) | (53.906) |
| hazard_event_count_financial | 23.125*** | 23.008** | 31.305 |
| | (8.011) | (9.886) | (29.456) |
| hazard_event_count_natural | -14.703* | 3.995 | -80.746*** |
| | (8.163) | (10.428) | (29.403) |
| hazard_event_count_predation | -53.416*** | -7.055 | -103.617 |
| | (13.783) | (15.181) | (63.404) |
| hazard_event_count_social | 3.792 | -0.457 | 34.661 |
| | (10.714) | (12.505) | (39.326) |
| nationalism_level | 4.046*** | 3.985*** | 3.940*** |
| | (0.004) | (0.009) | (0.007) |
| economic_conservativism | | | |
| social_conservativism | | | |
| anthropomorphic_promiscuity | | | |
| sociographic_prudery | | | |
| Constant | 1,672.924*** | -219.430 | 2,530.504*** |
| | (157.555) | (197.135) | (495.628) |
| Observations | 20,000 | 5,000 | 5,000 |
| R² | 0.986 | 0.979 | 0.989 |
| Adjusted R² | 0.986 | 0.979 | 0.989 |
| Residual Std. Error | 2,701.237 (df = 19975) | 1,571.367 (df = 4975) | 4,338.727 (df = 4975) |
| F Statistic | 57,688.900*** (df = 24; 19975) | 9,728.565*** (df = 24; 4975) | 18,118.520*** (df = 24; 4975) |

Note: *p<0.1; **p<0.05; ***p<0.01

To further investigate these relationships, we graphed the two nationalism clusters (high and low) and their effects on anti-immigrant sentiment as a function of how intense the different threat clusters are.

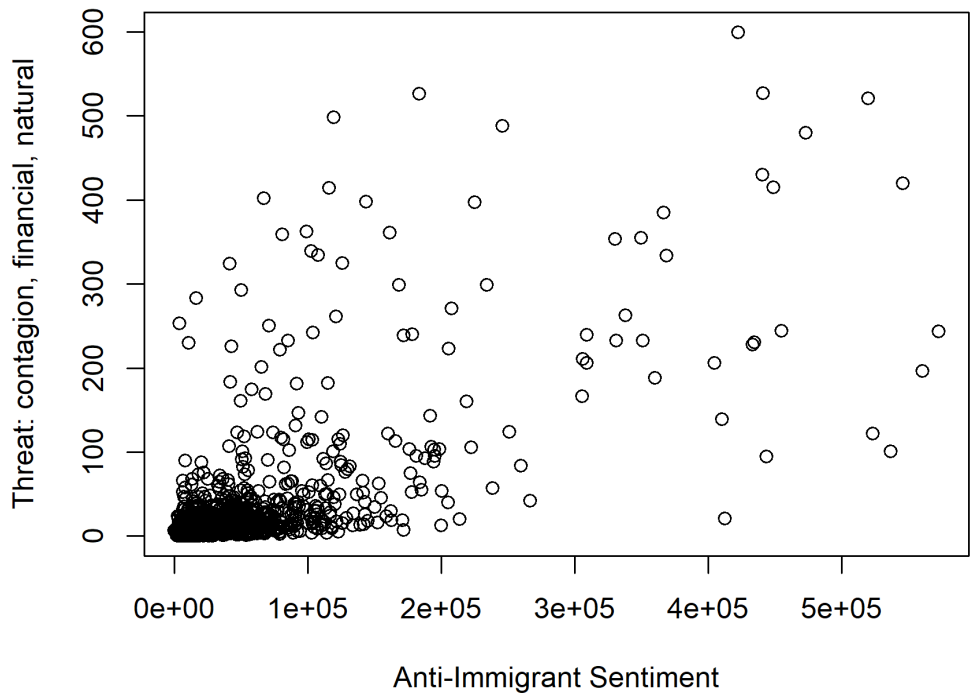

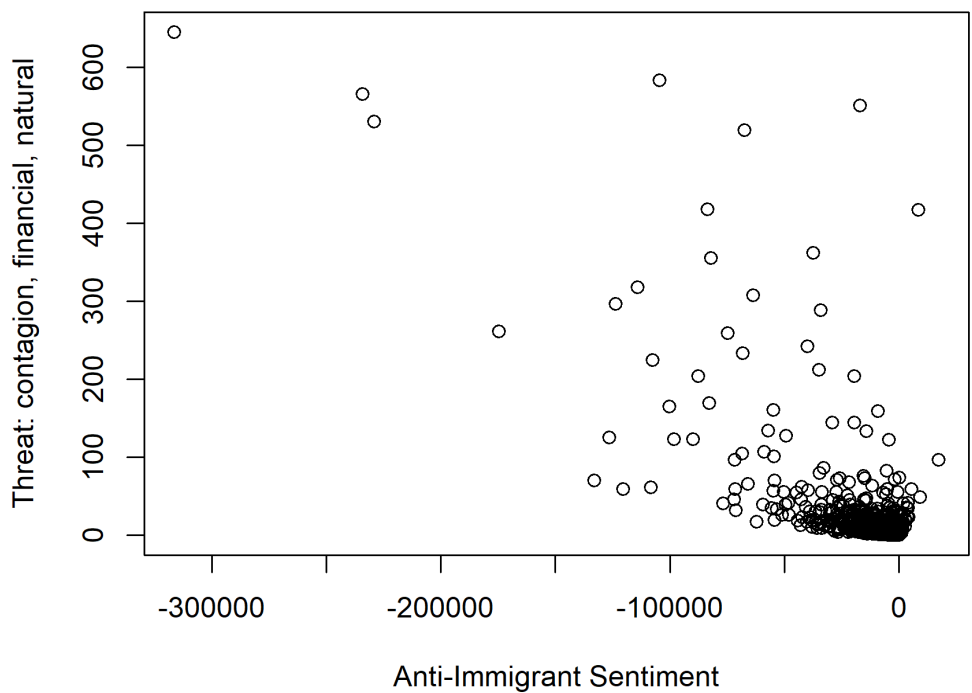

What is notable here is that there is a wide range of results, but very clear clusters for the two settings. When visualizing the same dynamics for social and predatory threats, we observed a similar pattern.

In addition, there is a correlation between media use and anti-immigrant values, but its only significant when religiosity is low. This suggests that religious values are guiding anti-immigrant values in some (but not all) circumstances. The data also seem to suggest that secular ideologies can also have an effect although that effect is different when looking at religious beliefs vs. religious identities.

In addition, those simulations resulting in high anthropomorphic promiscuity (likelihood of attributing things to supernatural agents) are not as affected by the media. But for those who have low anthropomorphic promiscuity, there is a significant negative effect on anti-immigrant sentiment from the percentage of the media that is negative.

In the table below, statistical results for all 4 regressions are presented in full. The data is subset as follows: model 1-low sociographic prudery, model 2-high sociographic prudery, model 3, low anthropomorphic promiscuity, model 4, high anthropomorphic promiscuity.

**Simulated data and anti-immigrant relationships for high and low religiosity variables**

| | Dependent variable: | | | |
|---|---|---|---|---|
| | anti_immigrant_sentiment | | | |
| | (1) | (2) | (3) | (4) |
| tvMediaUse | 199.861* | 311.240 | 248.818 | -54.863 |
| | (121.233) | (336.524) | (154.082) | (235.248) |
| threatPctOfMedia | -902.908*** | 462.309 | -468.036** | 76.086 |
| | (186.989) | (461.925) | (215.453) | (334.388) |
| socialMediaUse | -748.730*** | -52.424 | -293.661 | -128.884 |
| | (161.135) | (401.519) | (192.826) | (302.931) |
| rel_frequency | 291.233*** | 766.765*** | 254.587*** | 295.118*** |
| | (17.281) | (40.094) | (24.954) | (39.046) |
| initial_concern_social | 69.218 | -186.211 | -49.799 | -13.979 |
| | (77.618) | (215.516) | (99.004) | (154.437) |
| initial_concern_financial | -91.998 | 194.417 | -7.323 | -215.752 |
| | (77.476) | (213.138) | (98.685) | (153.704) |
| initial_concern_contagion | 246.201*** | 364.315* | 160.448* | 140.938 |
| | (77.158) | (214.463) | (97.164) | (154.590) |
| initial_concern_predation | -18.941 | -15.237 | 108.709 | -155.904 |
| | (77.017) | (212.145) | (98.501) | (154.998) |
| hazard_intensity_contagion | -1,859.861*** | -2,506.320*** | -1,550.441*** | -1,540.282*** |
| | (103.222) | (239.938) | (99.830) | (154.103) |
| hazard_intensity_financial | -1,515.354*** | -2,891.300*** | -1,313.597*** | -1,442.491*** |
| | (84.968) | (222.588) | (97.996) | (155.483) |
| hazard_intensity_natural | -930.353*** | -1,638.606*** | -834.133*** | -971.909*** |
| | (88.660) | (219.454) | (98.640) | (152.886) |
| hazard_intensity_predation | 1,668.852*** | 3,235.817*** | 1,457.795*** | 1,552.327*** |
| | (97.268) | (237.116) | (99.711) | (154.341) |
| hazard_intensity_social | 1,046.521*** | 2,969.025*** | 1,064.235*** | 1,114.912*** |
| | (82.051) | (220.674) | (99.368) | (156.227) |
| habituationRate | 1,626.501*** | -12,915.230*** | -858.049*** | -5,332.460*** |
| | (81.172) | (331.911) | (98.017) | (160.711) |
| energyDecay | 12.673 | 157.790 | 29.151 | 84.254 |
| | (76.883) | (211.905) | (97.860) | (154.014) |
| big_5_openness | 820.899*** | 690.610*** | 874.823*** | 420.123 |
| | (78.042) | (213.052) | (174.251) | (273.800) |
| big_5_conscientiousness | -288.639*** | -668.946*** | -120.307 | -674.821*** |
| | (78.188) | (210.549) | (117.023) | (182.319) |
| big_5_agreeableness | -341.067*** | -150.314 | -343.592*** | -0.670 |
| | (77.305) | (212.707) | (114.822) | (185.620) |
| hazard_event_count_contagion | -12.868 | -204.400*** | -37.993** | -65.732** |
| | (12.095) | (53.906) | (17.050) | (26.193) |
| hazard_event_count_financial | 23.008** | 31.305 | 23.715** | 36.398* |
| | (9.886) | (29.456) | (12.078) | (18.689) |
| hazard_event_count_natural | 3.995 | -80.746*** | -14.701 | -11.328 |
| | (10.428) | (29.403) | (12.143) | (19.179) |
| hazard_event_count_predation | -7.055 | -103.617 | -17.203 | -101.267*** |
| | (15.181) | (63.404) | (21.026) | (31.740) |
| hazard_event_count_social | -0.457 | 34.661 | 2.741 | -2.273 |
| | (12.505) | (39.326) | (15.891) | (24.679) |
| nationalism_level | 3.985*** | 3.940*** | 4.126*** | 3.986*** |
| | (0.009) | (0.007) | (0.009) | (0.007) |
| economic_conservativism | | | | |
| social_conservativism | | | | |
| anthropomorphic_promiscuity | | | | |
| sociographic_prudery | | | | |
| Constant | -219.430 | 2,530.504*** | 448.251* | 3,182.368*** |
| | (197.135) | (495.628) | (239.301) | (451.316) |
| Observations | 5,000 | 5,000 | 5,000 | 5,000 |
| $R^2$ | 0.979 | 0.989 | 0.982 | 0.987 |
| Adjusted $R^2$ | 0.979 | 0.989 | 0.982 | 0.987 |
| Residual Std. Error (df = 4975) | 1,571.367 | 4,338.727 | 1,999.826 | 3,143.568 |
| F Statistic (df = 24; 4975) | 9,728.565*** | 18,118.520*** | 11,276.280*** | 15,280.560*** |

Note: *p<0.1; **p<0.05; ***p<0.01

**Key Findings**

The goal of these studies was to assess complex questions during the COVID-19 pandemic around the relationships between nationalism, religiosity, and anti-immigrant sentiment from a socio-cognitive perspective. The intent of the project was to develop and test a system dynamics model that theoretically integrated and investigated these relationships, serving as a foundation for future research on the destabilizing effects of contagion threats. The importance of these issues is evident from daily news about the struggle between relatively liberal pan-European institutions and neo-nationalist parties. Many of the latter portray religion as an important source of their values, which shapes our increasingly socially and politically polarized environment.

Our simulation results suggest that the negative effect of the media and social media on anti-immigrant values is highest when religiosity is low, not when religiosity is high. So, while there is a correlation there between media and anti-immigrant values, it is only significant for people with lower religiosity. This indicates that religious values are guiding anti-immigrant values in some (but not all) circumstances.

Another important finding was that when nationalism is low, there is a significant effect of financial threats on anti-immigrant sentiment, but when nationalism is high there is no significant effect of financial threats on anti-immigrant sentiment. As noted above, this suggests that there is need for more research on the effect of the specific beliefs of a group on its members' acceptance of immigrants, and on the way in which individuals' framing of their national identity informs the extent to which they'll accept immigrants.

In assessing our threat measures, one of the things that stood out was that, as a scale, our threat measure did not have good validity as a single measure. We found that contagion and financial threats have a significant negative relationship with immigrant attitudes, while social threats have a positive relationship. This suggests that those who perceive more social threats are less likely to want immigrant neighbors, but those who perceive more contagion or financial threats are less likely to respond negatively to questions about having immigrant neighbors.

**Directions for Further Research**

The survey data leaves several unanswered questions for future research. First, there is a well-documented effect of moral values being correlated with different political persuasions, as discussed briefly in the theoretical background section above. The pattern discovered here between the different threats suggests that evolved human responses to environmental threats and contexts also help to explain social and economic beliefs. It is possible that threats are an underlying cause for the moral domains that are observed in the literature. Future research needs to investigate the relationship between threat and moral domains as causal forces in political ideologies. Ideally, this would be done using controlled or quasi-experimental methods in order to better untangle the possible causal directions.

The relationships among different styles of identity fusion and nationalism and religiosity should also be investigated. In particular, more careful, structured, qualitative research could help to uncover the role of beliefs and semiotically distinct ways of self-identification as potentially important causal factors shaping support for immigration policies. The results of the survey were consistent with earlier literature demonstrating the extreme "groupishness" that exists for fused individuals, as discussed briefly above, but also shed light on previously undocumented effects of media use and content on fusion. Future research can utilize previously deployed methods for analysing fusion as it pertains to ingroup beliefs, which can be inferred from texts and qualitative fieldwork interviews.

The survey data suggested an interesting correlation between religiosity and nationalism. This should be further investigated as both appear to be correlated with anti-immigrant values, although the relationship can be moderated depending on religious group. This indicates that there is a critical role of religious beliefs in that relationship, and the extent to which they are informed by environmental threats is an open question.

The data from the simulation also provide a foundation for further research. For example, the simulated data brought to light a very strong negative relationship between openness as a personality trait and anthropomorphic promiscuity. Previous research has shown that some personality variables do correlate with different patterns of belief/non-belief and religious affiliation. However, the extent to which these affect specific dimensions of religiosity, such as anthropomorphic promiscuity and sociographic prudery, is still underexplored.

Both the survey and the simulation suggest that nationalism and religion are affected by the same variables. As such, religion might not be a cause of nationalism (or nationalism the cause of religion), but they could be correlated because of mutual causation. Generally, however, there is a clear trend where more frequent religious attendance can result in higher levels of nationalism. This is probably the most pressing unanswered question of this research. If in fact religion and nationalism are correlated, it could be because they both have some of the same underlying causes. It could be that the same underlying cognitive capacities or tendencies are the foundation for both religion and nationalism, but that religion and nationalism per se have independent effects on anti-immigrant sentiment. If so, it would be problematic to assume that the religious and nationalist effects on anti-immigrant support should be addressed in exactly the same way.